\documentclass[prb,preprint,showpacs,preprintnumbers,amsmath,amssymb,floatfix]{revtex4}

\usepackage{graphicx}% Include figure files
\usepackage{dcolumn}% Align table columns on decimal point
\usepackage{bm}% bold math

\begin{document}

%\preprint{APS/123-QED}

\title{Dynamic surface critical behavior of isotropic Heisenberg
    ferromagnets: boundary conditions, renormalized field theory, and
    computer simulation results}% Force line breaks with \\

\author{H.\ W. Diehl}
\author{M.\ Krech}
\altaffiliation[Present address: ]{Institut f\"ur Theoretische und
Angewandte Physik, Universit\"at Stuttgart, 70550 Stuttgart and
Max-Planck-Institut f\"ur Metallforschung, Heisenbergstr. 1, 70569 Stuttgart,
Federal Republic of Germany}
\author{H.\ Karl}
\affiliation{%
Fachbereich Physik, Universit{\"a}t Essen, 45117 Essen,
    Federal Republic of Germany
}%

\date{\today}

\begin{abstract}
  The dynamic critical behavior of isotropic Heisenberg ferromagnets
  with a planar free surface is investigated by means of
  field-theoretic renormalization group techniques and high-precision
  computer simulations. An appropriate semi-infinite extension of the
  stochastic model J is constructed. The relevant boundary terms of
  the action of the associated dynamic field theory are identified,
  the implied boundary conditions are derived, and the renormalization
  of the model in $d<6$ bulk dimensions is clarified. Two distinct
  renormalization schemes are utilized. The first is a massless one
  based on minimal subtraction of dimensional poles and the
  dimensionality expansion about $d=6$. To overcome its problems in
  going below $d=4$ dimensions, a massive one for fixed dimensions
  $d\le 4$ is constructed. The resulting renormalization group (or
  Callan Symanzik) equations are exploited to obtain the scaling forms
  of surface quantities like the dynamic structure factor. In
  conjunction with boundary operator expansions scaling relations
  follow that relate the critical indices of the dynamic and static
  infrared singularities of surface quantities to familiar
  \emph{static} bulk and surface exponents. To test the predicted
  scaling forms and scaling-law expressions for the critical exponents
  involved, accurate computer-simulation data are presented for the
  dynamic surface structure factor. These are in conformity
  with our predictions.
\end{abstract}
\pacs{75.10.Hk, 68.35.Rh, 64.60.Ht, 05.70.Jk}
%\keywords{Suggested keywords}%Use showkeys class option if keyword
                              %display desired
\maketitle

\section{Introduction}
\label{sec:Intro}

A cornerstone of the modern theory of critical phenomena is the
arrangement of microscopically different systems in universality
classes of equivalent critical behavior.\cite{MEF98,Fis83} Few basic
properties, such as the spatial dimension $d$, the order-parameter
symmetry, and gross features of the interactions determine to which
universality class for static bulk critical behavior a particular
system belongs. These universality classes can be represented by
simple continuum models like the $\phi^4$ model, which are minimal in
the sense that dropping any of the Hamiltonian's terms implies a
change of the universality class. An important alternative way of
representing the universality classes is through standard lattice
(spin) models such as the Ising model, which lend themselves best for
precise Monte Carlo simulations.

A similar classification scheme exists for \emph{dynamic} bulk
critical behavior.\cite{HH77} The associated
universality classes---called
dynamic bulk universality classes henceforth---%
additionally depend on basic properties of the dynamics such as
conservation laws, and since distinct dynamics may have the same
equilibrium distribution, each \emph{static} universality class
generally splits up into \emph{several dynamic} ones. The latter are
represented by stochastic models called A, B, \ldots, J.\cite{HH77}

Research over the past 25 years has revealed the existence of
analogous universality classes for \emph{static surface} critical
behavior of semi-infinite systems at \emph{bulk critical
  points}.\cite{Bin83,Diehl} To which static surface universality
class a given system belongs is decided by its static bulk
universality class and additional relevant surface properties.  Hence
each static surface universality class as well as each dynamic bulk
universality class usually splits up into separate dynamic surface
universality classes. Furthermore, systems belonging to the same
\emph{static surface} universality class and the same \emph{dynamic
  bulk} universality class may be representative of \emph{distinct
  dynamic surface} universality classes as local changes of the
dynamics at the surface can be relevant.\cite{ModB_A,DJ92,Die94b,WD95}

Unfortunately, the number of detailed theoretical investigations of
dynamic surface critical behavior performed until now is rather
limited.\cite{DJ92,Die94b,WD95,DD83b,FD,XiGo} Furthermore, they
focused more or less exclusively on models with purely
\emph{relaxational dynamics}.  On the experimental side, the situation
is worse: stringent experimental checks of the theoretical predictions
for dynamic surface critical behavior, though urgently needed, are
still lacking. One obvious reason for this is the difficulty of such
experiments. The impressive progress made during the past two decades
in the perfection of surface-sensitive scattering techniques has so
far led only to accurate experimental investigations of \emph{static}
surface critical behavior.\cite{MDPJ90,Dos92,KDN+97,Xraytheory}
Demonstrating that similarly conclusive data can also be obtained for
\emph{dynamic} surface critical behavior remains a major experimental
challenge, albeit such experiments are expected to become feasible in
the near future. According to the recent TESLA design
report,\cite{fel} the X-ray free electron laser offers a great
potential for such experiments.

Theoretical progress can play an important role in stimulating such
experiments. We believe that theoretical advances in two directions
are essential for achieving this goal. On the one hand, models
representing other bulk dynamic universality classes must be
considered, generalized to systems with boundaries, and carefully
investigated to find out what kinds of dynamic surface critical
behavior can occur, i.e., which dynamic surface universality classes
exist. On the other hand, detailed theoretical predictions should be
worked out for \emph{experimentally accessible quantities} like
structure functions etc.

Pursuing these goals, we will investigate the dynamic surface critical
behavior of \emph{isotropic Heisenberg ferromagnets} in this paper.
Well-known characteristic features of the dynamics of such magnets are
the presence of \emph{nondissipative (mode-coupling) terms} and the
\emph{conservation of the order parameter}.  We shall employ two
different lines of approaches: (i) analytic work based on the
field-theoretic renormalization group (RG) and (ii)
computer-simulation studies of the dynamic surface structure function.
A brief account of parts of our work has been given
elsewhere.\cite{KKD01}

In our RG work we utilize an appropriate semi-infinite extension
of the usual stochastic bulk model
J,\cite{HH77,MM75,Kaw75,Jan76,BJW76,Doh76,Jan79} which represents the
dynamic bulk universality class of the isotropic Heisenberg
ferromagnet, without energy conservation. A familiar problem one is
faced with is the following. Whereas the upper critical dimension of
this dynamic model is $d^*_{\text{J}}=6$, the one of its
steady-state distribution, described by the usual
$|{\bm{\phi}}|^4$ model with an $n=3$ vector field
${\bm{\phi}}$, is $d^*=4$. Thus the small
parameter in which a dimensionality expansion can be made in the
dynamic case is $\epsilon_6=6-d$ rather than $\epsilon_4=4-d$, where
$d$ is the bulk dimension. For $4<d\le 6$, the static critical
behavior is given by mean-field theory and associated with the (then
infrared-stable) Gaussian fixed point of the $|{\bm{\phi}}|^4$
theory, even though the dynamic critical behavior is described by a
nontrivial fixed point that is characterized by a non-zero value $f^*$
of the mode-coupling vertex and accessible to the $\epsilon_6$
expansion.

Unfortunately, this expansion is \emph{not tailored} to capture the
nontrivial static critical exponents that emerge as $d$ drops below
$4$. Therefore it is of somewhat limited use in the physically
interesting three-dimensional case or, more generally, for $d\le 4$.
In order to to find out which scaling laws exist relating the critical
exponents of dynamic bulk and surface quantities to known bulk and
surface critical indices, it is essential to formulate the
field-theoretic RG for fixed values of $d<4$. We do this by
\emph{extending existing massive RG schemes for semi-infinite systems
  \cite{DS94,DS98} to dynamics}. This yields RG (Callan-Symanzik)
equations whose exploitation in conjunction with known boundary
operator expansions\cite{Diehl,DD81c,MO93} reveals that the dynamic
bulk and surface critical exponents can be expressed completely in
terms of known static ones, besides giving the scaling forms of
quantities like the surface structure function.

In order to check these finding we have performed high-precision
computer simulations of a semi-infinite lattice model of classical
Heisenberg spins whose dynamics is defined via the deterministic
nondissipative equations of motion implied by their Poisson bracket
relations. The advantage of this simple dynamics without noise is that
recently developed extremely efficient spin dynamics algorithms
\cite{SD,DPLMK} can be employed to compute the temporal development
from given initial spin configurations, which we choose from a thermal
equilibrium distribution generated via a Monte Carlo simulation.

It should be emphasized that this lattice model differs in an
important aspect from the continuum model we consider: unlike the
latter, it \emph{conserves the energy}. Nevertheless, both models
belong to the same universality class, as we intend to show.

The remainder of this paper is organized as follows. In
Sec.~\ref{sec:mod} both the semi-infinite lattice model (studied by
simulations) as well as the semi-infinite extension of the continuum
model J (utilized in our RG analysis) are introduced, and their
dynamics specified. The definition of the continuum model also
involves the specification of appropriate boundary conditions. We
discuss this question first on a heuristic basis (Sec.~\ref{sec:bcs}).
Going over to the path-integral formulation of this model in
Sec.~\ref{sec:FIF}, we then show in Sec.~\ref{sec:bcphitilde} how the
boundary conditions for both the order parameter $\bm{\phi}$ and the
auxiliary (Martin-Siggia-Rose) field $\tilde{\bm{\phi}}$ can be
justified in a systematic manner and derived from the boundary part of
the dynamic action functional. Section \ref{sec:fdt} briefly recalls
the fluctuation-dissipation theorem and discusses the meaning of some
of the boundary conditions in this context. Section \ref{sec:RG} is
devoted to the RG analysis of the continuum model. After giving
the free response and correlation propagators in
Sec.~\ref{sec:prelim}, we explicate in Sec.~\ref{sec:RS1} the
renormalization of the theory, describe the massless renormalization
scheme on which our subsequent RG analysis in $6-\epsilon_6$
dimensions is based. To overcome the limitations of this scheme, we
construct in Sec.~\ref{sec:RS2} a massive RG scheme for fixed
dimensions $d$ with $2<d\le 4$. The resulting Callan-Symanzik
equations are given in Sec.~\ref{sec:CSE} and utilized to derive the
scaling forms of the correlation and response functions. Details of
our Monte-Carlo spin dynamics simulation are described in
Sec.~\ref{sec:MC}. Its results are presented and analyzed in
Sec.~\ref{sec:dynssf}. Section \ref{sec:concl} contains a brief
summary and concluding remarks. Finally, in the Appendix arguments are
given as to why the lattice model we study belongs to the
universality class of our semi-infinite model J, even though it
differs from the latter by conserving additionally energy.

\section{Models}
\label{sec:mod}

\subsection{Semi-infinite lattice Heisenberg model}
\label{sec:lHm}

The lattice model we consider is a classical isotropic Heisenberg
ferromagnet on a $d$-dimensional simple cubic lattice whose sites
$\bm{i}=(i_1,\ldots,i_d)$, with $i_\kappa=0,\ldots,L-1$
for $\kappa=1,\ldots,d$, are occupied by spins
$\bm{S}_{\bm{i}}=(S^{\alpha}_{\bm{i}},\,\alpha{=}1,2,3)$
of length $|\bm{S}_{\bm{i}}|=1$. Free boundary
conditions apply along the $i_d$-direction and periodic ones along the
remaining $d-1$ ones, so that the layers $i_d=0$ and $i_d=L-1$ are
free surfaces. The Hamiltonian of the model reads
\begin{equation}
\label{eq:Hamil}
{\mathcal{H}}_{\text{lat}} = -J
\sum_{\stackrel{\langle \bm{i},\bm{j} \rangle}{i_d\,\text{or}\,j_d\ne 0,L-1}}
{\bm{S}}_{\bm{i}} \cdot {\bm{S}}_{\bm{j}}
- J_1 \sum_{\stackrel{\langle \bm{i},\bm{j} \rangle}{i_d=j_d = 0,L-1}}
{\bm{S}}_{\bm{i}}\cdot{\bm{S}}_{\bm{j}}\;,
\end{equation}
where the summations run over the specified sets of nearest-neighbor
(nn) bonds $\langle \bm{i},\bm{j} \rangle$. The bulk
and surface nn interaction constants $J$ and $J_1$ are ferromagnetic
and measured in temperature units $k_{\text{B}}T$. The dynamics is
defined by the equations of motion
\begin{equation}
\label{eq:eqmot}
\frac{d \bm{S}_{\bm{i}}}{dt} =
\frac{\partial {\mathcal H}_{\text{lat}}}{\partial \bm{S}_{\bm{i}}}
\times {\bm{S}}_{\bm{i}}\;, 
\end{equation}
which describe the precession of the spins ${\bm{S}}_{\bm{i}}$ in the
local magnetic fields ${\bm{H}}_{\bm{i}}\propto -\partial{\mathcal
  {H}}_{\text{lat}}/ \partial {\bm{S}}_{\bm{i}}$. They conserve both
total spin $\sum_{\bm{i}}{\bm{S}}_{\bm{i}}$ (in the here assumed
absence of external magnetic fields) as well as total energy
$E_{\text{lat}}\equiv {\mathcal{H}}_{\text{lat}}[{\bm{S}}(t)]$.
  
Conservation of magnetic energy is not normally considered a property
of real ferromagnets since the spin system can loose energy by
processes not taken into account by Eqs.~(\ref{eq:Hamil}) and
(\ref{eq:eqmot}) such as coupling to phonons. In fact, in the
continuum model J employed in our RG analysis, only the order
parameter but not the energy is conserved. Arguments as to why both
models represent nevertheless the same universality class are given in
Appendix \ref{sec:ec}.

In our computer simulations, a $d{=3}$ dimensional version of the above
model is investigated. The equations of motion (\ref{eq:eqmot}) are
numerically integrated for a given set of at least $700$ initial spin
configurations generated by a Monte Carlo simulation of the thermal
equilibrium state associated with Hamiltonian
(\ref{eq:Hamil}).\cite{SD,DPLMK,MKslab} Details of this simulation are
explained in Sec.~\ref{sec:MC}.

Quantities of primary importance for the interpretation of scattering
experiments are the spin-spin cumulant
\begin{eqnarray}
  \label{eq:Sijtt}
&&C^{\alpha \beta}{\left(\bm{r}; z, z'; t {-} t' \right)}
\equiv {\big\langle S_{\bm{i}}^{\alpha}(t)\,
S_{\bm{i}'}^{\beta}(t')  \big\rangle}^{\text{cum}}\nonumber\\
&&={\big\langle S_{\bm{i}}^{\alpha}(t)\,
S_{\bm{i}'}^{\beta}(t')  \big\rangle}-{\big\langle S_{\bm{i}}^{\alpha}(t)\big\rangle}{\big\langle
S_{\bm{i}'}^{\beta}(t')  \big\rangle}
\end{eqnarray}
and its Fourier transform 
\begin{eqnarray}
  \label{eq:FTC}
 && \hat{C}^{\alpha \beta}(\bm{p};z,z';\omega)\nonumber\\
 && ={\int}d^{d-1}r\,e^{-i\,\bm{p}\cdot \bm{r}}
 {\int_{-\infty}^\infty}dt\,e^{i\omega t}\,C^{\alpha\beta}
 {\left(\bm{r}; z, z'; t {-} t' \right)}\;.
\end{eqnarray}
Here $\bm{r}=(i_1{-}i_1',\ldots,i_{d-1}{-}i'_{d-1})$, while
$z=i_3$ and $z'=i_3'$, respectively. Further, $t$ and $t'$ are times
to which the initial spin configuration at $t=0$ has evolved according
to Eq.~(\ref{eq:eqmot}). The average $\langle{.}\rangle$ is taken
over the distribution of initial configurations.

Specifically, we will be concerned with the dynamic surface structure
function
\begin{equation}
  \label{eq:C11pomega}
  {\hat{C}^{\alpha \beta}_{11}}(\bm{p},\omega)\equiv
  \hat{C}^{\alpha \beta}(\bm{p};0,0;\omega)\;.
\end{equation}
Before embarking on a discussion of its scaling properties and
presenting our simulation results, it is useful to introduce first the
continuum model on which our RG analysis is based.

\subsection{Semi-infinite model J}
\label{sec:simJ}

\subsubsection{Hamiltonian of the thermal equilibrium state}
\label{sec:contHam}

The dynamic model we are going to consider is required
to satisfy detailed balance \cite{HH77,DJ92,Jan92} and to
ensure relaxation to a steady-state distribution corresponding to a
thermal equilibrium state $\propto e^{-{\mathcal H}[\bm{\phi}]}$
with the Hamiltonian
\begin{equation}
\label{eq:HGL}
{\mathcal H} = {\int_{{\mathbb{R}}^d_+}} \left[\frac{1}{2}\,
(\bm{\nabla} \bm{\phi})^2 + \frac{\tau_0}{2}\,\phi^2 +
\frac{u_0}{4!}\,|\bm{\phi}|^4\right] + {\int_{\mathcal B}}
\frac{c_0}{2}\, \phi^2 \;.
\end{equation}
Here the integrations extend over
${\mathbb{R}}^d_+\equiv\{{(\bm{x}_{\|},z)}\in{\mathbb{R}}^d\mid
z\ge 0\}$, the $d$-dimensional half-space, and ${\mathcal B}$, its
($d-1$)-dimensional boundary plane at $z=0$, respectively. The
order-parameter density ${\bm{\phi}}=(\phi^\alpha)$ is a
three-vector.

Above $d=3$ bulk dimensions, this static model is known to undergo at
the bulk critical point so-called ordinary, special, and extraordinary
surface transitions, depending on whether the surface enhancement
variable $c_0$ is larger than, equal to, or less than a critical value
$c_{\text{sp}}$.\cite{Bin83,Diehl} For $d=3$, the surface cannot
spontaneously order at the bulk critical temperature $T_{\text{c}}>0$
because of the presumed continuous $O(3)$ symmetry of the Hamiltonian
(\ref{eq:HGL}). Hence only the \emph{ordinary} transition remains in
this case. Analogous statements apply to the lattice model
(\ref{eq:Hamil}), for whose $d>3$ variant the role of the variable
$-c_0$ is played by the `surface enhancement'
$(J_1/J)-(J_1/J)_{\text{sp}}$, where $(J_1/J)_{\text{sp}}$ is the
critical value of the ratio $J_1/J$ pertaining to the special
transition.

\subsubsection{Langevin equations}
\label{sec:Langeq}

Next we turn to the task of formulating an appropriate semi-infinite
extension of the standard bulk model J. For reasons expounded
elsewhere,\cite{DJ92,Die94b} we may assume that the surface-induced
modifications of both the interactions as well as the dynamics are
restricted to the immediate vicinity of the boundary ${\mathcal{B}}$.
Consequently, we use the stochastic bulk equation
\begin{equation}\label{eq:stocheqJ}
\dot{\bm{\phi}}(\bm{x},t)
=\lambda_0\,{\left(\triangle\,
{\mathcal{H}}_{\bm{\phi}}
+f_0\,{\mathcal{H}}_{\bm{\phi}}
\times{\bm{\phi}}\right)}+\bm{\zeta}(\bm{x},t)
\end{equation}
for all points ${\bm{x}}$ with $z>0$.
Here $\bm{\zeta}$ is a Gaussian random force
with average $\langle\bm{\zeta}\rangle=0$ and variance
\begin{equation}\label{eq:varzeta}
{\left\langle{\zeta^\alpha}(\bm{x},t)\,
{\zeta^\beta}(\bm{x}',t')
\right\rangle}=-2\lambda_0\,{\delta^{\alpha\beta}}%\,
\triangle\,\delta(\bm{x}{-}\bm{x}')\,\delta(t{-}t')\;.
\end{equation}
Further, ${\mathcal{H}}_{\bm{\phi}}$ stands for the part of
the functional derivative
\begin{equation}
  \label{eq:delHdelphi}
  \frac{\delta{\mathcal H}}{\delta\bm{\phi}}(\bm{x},t)=
  {\mathcal{H}}_{\bm{\phi}}(\bm{x},t)
  +\delta(z)\,(c_0-\partial_n)\,\bm{\phi}(\bm{x},t)
\end{equation}
that remains away from the boundary plane ${\mathcal B}$, namely
\begin{equation}
  \label{eq:Hphi}
  {\mathcal{H}}_{\bm{\phi}}(\bm{x},t)=
{\left(-\triangle+\tau_0+\frac{u_0}{6}\,|\bm{\phi}|^2\right)}\bm{\phi}\;.
\end{equation}
The derivative $\partial_n$ in Eq.~(\ref{eq:delHdelphi}) is along the
inner normal, i.e., $\partial_n=\partial_z$ on ${\mathcal B}$.

In order to extend the model to the semi-infinite case, we must
specify whether and how Eqs.~(\ref{eq:stocheqJ}) and
(\ref{eq:varzeta}) are to be modified in the vicinity of ${\mathcal
  B}$.  Owing to our locality assumption mentioned at the beginning of
this sub-subsection, this should amount to a choice of boundary
conditions for $\bm{\phi}$. For the sake of simplicity, we assume that
the conservation of the order parameter is \emph{not} violated by
boundary contributions. This is physically reasonable since we took
all bulk \emph{and} surface terms of the Hamiltonian (\ref{eq:HGL}) to
have $O(3)$ symmetry, as is appropriate for a Heisenberg magnet whose
interactions are isotropic even at the surface.

\subsubsection{Boundary conditions for $\bm{\phi}$}
\label{sec:bcs}

Building on previous work on model B\cite{DJ92,Die94b,WD95}, we can
now easily anticipate the proper boundary conditions. One boundary
condition for $\bm{\phi}$ should be the usual static one 
\begin{equation}
  \label{eq:stbcphi}
  \partial_n\bm{\phi}=c_0\,\bm{\phi}\;,
\end{equation}
which ensures the vanishing of the  contribution $\propto\delta(z)$ of
the functional derivative (\ref{eq:delHdelphi}).

The other one is entailed by the required order-parameter
conservation. This becomes clear if we rewrite Eq.~(\ref{eq:stocheqJ})
as a continuity equation: 
\begin{equation}
  \label{eq:conteq}
  \dot{\phi^{\alpha}}+\bm{\nabla}\cdot {\left(\bm{j}^{(\alpha)}+\bm{j}_{\bm{\zeta}}^{(\alpha)}\right)}=0\;.
\end{equation}
Here
\begin{equation}
  \label{eq:j}
  \bm{j}^{(\alpha)}=-\lambda_0\,{\left(\bm{\nabla}{\mathcal{H}}_{\phi^\alpha}+f_0\,\epsilon^{\alpha\beta\gamma}\phi^\beta\bm{\nabla}\phi^\gamma\right)}
\end{equation}
are the deterministic parts of the currents, and the noise parts satisfy
\begin{equation}
  \label{eq:jzeta}
 \zeta^\alpha= -\bm{\nabla}\cdot\bm{j}_{\bm{\zeta}}^{(\alpha)}\;.
\end{equation}

To ensure conservation of the total order parameter, no currents must
leave the system. Hence the normal component of the currents should
vanish,
\begin{equation}
  \label{eq:curcons}
  j^{(\alpha)}_n\equiv\bm{n}\cdot\bm{j}^{(\alpha)}=0\;,\quad\alpha=1,2,3.
\end{equation}
If spin anisotropies were present at the surface (which is not
uncommon), the conservation would be violated at the surface for some,
if not all, components of $\bm{\phi}$.

Both boundary conditions, Eqs.~(\ref{eq:stbcphi}) and
(\ref{eq:curcons}), are valid in an operator sense, i.e., hold inside
of averages over the initial values and the noise (yielding
correlation and response functions). Note that the validity of the former
has two immediate consequences: The surface contributions
to the currents one would expect from the $\delta$-function term of
Eq.~(\ref{eq:delHdelphi}) upon using the definition
$\bm{j}^{(\alpha)}=-\lambda_0\,\bm{\nabla}\delta{\mathcal{H}}/\delta\phi^\alpha$
rather than Eq.~(\ref{eq:j}) disappears. Furthermore, substitution of
the boundary condition (\ref{eq:stbcphi}) into Eq.~(\ref{eq:curcons})
shows that the precession term's contributions ($\propto f_0$) to the
currents $\bm{j}^{(\alpha)}$, $\alpha=1,2,3$, vanish, so
that these latter boundary conditions become
\begin{equation}
  \label{eq:scc}
  \partial_n{\mathcal{H}}_{\bm{\phi}}\equiv\partial_n{\left(-\triangle+\tau_0+\frac{u_0}{6}\,|\bm{\phi}|^2\right)}\bm{\phi}=0\;.
\end{equation}

The probability distribution of the noise clearly must also comply
with the presumed order-parameter conservation. We prefer to discuss
the consequences within the framework of the functional-integral
(re-)formulation of the theory,\cite{Jan76,BJW76,Jan92,dDom76} where
they manifest themselves as boundary conditions for the auxiliary or
Martin-Siggia-Rose\cite{MSR73} (MSR) field $\tilde{\bm{\phi}}$ 
introduced below.

\subsubsection{Functional-integral formulation}
\label{sec:FIF}

The Langevin equations (\ref{eq:stocheqJ}) can be rewritten as
\begin{equation}
  \label{eq:Leqintermsro}
  \dot{\bm{\phi}}(\bm{x},t)=
-{\left({\bm{\mathcal R}}\cdot
\frac{\delta{\mathcal H}}{\delta\bm{\phi}}\right)}(\bm{x},t)+\bm{\zeta}(\bm{x},t)\:,
\end{equation}
where $\bm{\mathcal R}=({\mathcal R}^{\alpha\beta})$ denotes
the reaction operator
\begin{equation}
\label{eq:reactop}
{\mathcal R}^{\alpha\beta}=-\lambda_0\,{\left(\delta^{\alpha\beta}\triangle
+f_0\,{\epsilon^{\alpha\beta\gamma}}\,\phi^{\gamma}\right)}\;.
\end{equation}
Since this operator acts on ${\mathcal H}_{\bm{\phi}}$, which
according to Eq.~(\ref{eq:scc}) satisfies a Neumann boundary
condition, the Laplacian it involves is self-adjoint on an appropriate
space of (sufficiently smooth) functions satisfying this boundary
condition.

The measure $e^{-{\mathcal J}[\tilde{\bm{\phi}},\bm{\phi}]}
{\mathcal D}[\tilde{\bm{\phi}},\bm{\phi}]$ which appears in the
equivalent functional-integral formulation \cite{Jan76,BJW76,Jan92,dDom76}
of the theory can now easily be inferred. To this end, let us first recall
which form the action ${\mathcal J}[\tilde{\bm{\phi}},\bm{\phi}]$ must have
to ensure detailed balance and relaxation to the chosen equilibrium state.
For the here considered case in which the noise has a Gaussian probability
distribution, this is\cite{Jan76,Jan92,DJ92}
\begin{equation} \label{eq:Jdbf}
{\mathcal J} = {\int_{-\infty}^{\infty}}\!dt\,
{\int_{\bm{x}}}
{\left\{\tilde{\bm{\phi}}\cdot\!
{\left[\dot{\bm{\phi}}
+ \tensor{\bm{{\mathcal R}}}{\cdot}{\left(
\frac{\delta {\mathcal H}}{\delta \bm{\phi}}
-\tilde{\bm{\phi}}\right)}-
\frac{\delta\bm{\tensor{\mathcal R}}}{\delta\bm{\phi}}
\right]} \right\}}\;,
\end{equation}
where a pre-point discretization of time is understood to be employed.
 
The action of the bulk model J is known to be of this form, with the
reaction operator $\tensor{\mathcal R}$ being given by
Eq.~(\ref{eq:reactop}). If we accept the boundary conditions
(\ref{eq:stbcphi}) and (\ref{eq:scc}), then contributions to the
action that are localized on the surface vanish. Consequently this
result for the action must also hold in the semi-infinite case we
considered, with $\int_{\bm{x}}$ interpreted as the volume
integral $\int_{{\mathbb{R}}^d_+}$.

Conversely, one can start from an action of form (\ref{eq:Jdbf}) and
derive the boundary conditions in a systematic fashion\cite{Kar00}
along the lines followed in Refs.~\onlinecite{DJ92} and
\onlinecite{Die94b}. The various general assumptions we have made
(consistency with bulk model J, only local modifications of the
dynamics at the surface, absence of nonconservative surface terms, etc.)
can be combined with relevance/irrelevance considerations to conclude
that the reaction operator reads
\begin{equation}
\label{eq:symreactop}
\tensor{\bm{{\mathcal R}}}=\lambda_0\,{\left(\delta^{\alpha\beta}\,\loarrow{\bm{\nabla}}\roarrow{\bm{\nabla}}
-f_0\,{\epsilon^{\alpha\beta\gamma}}\,\phi^{\gamma}\right)}\;,
\end{equation}
where $\loarrow{\bm{\nabla}}$ acts to the left. This is
identical to Eq.~(\ref{eq:reactop}) up to the replacement of the
Laplacian by the symmetric expression
$\loarrow{\bm{\nabla}}\roarrow{\bm{\nabla}}$.

The substitution of this form of $\tensor{\bm{\mathcal R}}$ into
Eq.~(\ref{eq:Jdbf}) and an integration by parts (making no use of the
boundary conditions (\ref{eq:stbcphi}) and (\ref{eq:scc})) yields
\begin{eqnarray}
  \label{eq:JmodJ}
  {\mathcal J} &=& {\int_{-\infty}^{\infty}}\!dt\biggl(
{\int_{{\mathbb{R}}^d_+}}
{\Big\{}\tilde{\bm{\phi}}\cdot{\big(
\dot{\bm{\phi}}-\lambda_0\,f_0\,{\mathcal{H}}_{\bm{\phi}}
\times{\bm{\phi}}\big)}\nonumber\\
&&\mbox{}-\lambda_0\, {\big(
{\mathcal H}_{\bm{\phi}}
-\tilde{\bm{\phi}}\big)}\triangle\tilde{\bm{\phi}}
\Big \}
\nonumber\\
&&\mbox{}
-\lambda_0\int_{\mathcal{B}}\!{\Big\{}
(\triangle\tilde{\bm{\phi}})\cdot
{(c_0-\partial_n)} {\bm{\phi}}
+{\big[}
{\mathcal{H}}_{\bm{\phi}}-\tilde{\bm{\phi}}
\nonumber\\
&&\mbox{}
+\delta(z)\,{(c_0-\partial_n)} {\bm{\phi}}
-f_0\,{\bm{\phi}}\times\tilde{\bm{\phi}}{\big]}\!\cdot{
\partial_n}\tilde{\bm{\phi}}
\Big\}
{\biggr)}.
\end{eqnarray}
The singular piece $\propto \delta(z{=}0)$ present in the boundary
integral $\int_{{\mathcal B}}$ is familiar from
Refs.~\onlinecite{DJ92,Die94b,WD95}. It results from the
coincidence of two $\delta$ functions. This singularity can be cured
by replacing one of the $\delta$ functions by a smeared-out smooth
analog such as $\delta_B(z)\equiv B\,e^{-B\,z}$, with a large positive
but finite value of $B$.

\subsubsection{Boundary conditions as boundary equations of motion}
\label{sec:bcphitilde}

Starting from the action (\ref{eq:Jdbf}), we can now obtain the
boundary conditions for both $\bm{\phi}$ and $\tilde{\bm{\phi}}$ in a
standard manner\cite{DJ92,Die94b} as `boundary contributions to the
equations of motion'. This works as follows. We add source terms
${\mathcal J}_{j}[\tilde{\bm{\phi}},\bm{\phi}]$ to the action and
consider the generating functional ${\mathcal Z}[j]\equiv
{\int}{\mathcal D}[\tilde{\bm{\phi}},\bm{\phi}]\,e^{-{\mathcal
J}-{\mathcal J}_{j}}$, where $j$ stands for the set of all sources
considered, including eventual ones localized on the surface. That is,
the source part of the action can be written as
\begin{equation}
  \label{eq:srcts}
  {\mathcal J}_{j}[\tilde{\bm{\phi}},\bm{\phi}]
=-{\int_{-\infty}^\infty}dt{\left[
\int_{{\mathbb{R}}^d_+}j_\kappa\,O_\kappa
+\int_{\mathcal{B}}j_\rho\,O_\rho
\right]}\;,
\end{equation}
where $O_\kappa=O_\kappa({\bm{x}},t)$ and
$O_\rho=O_\rho({\bm{x}}_\|,t)$ are local functionals of
$\bm{\phi}$ and $\tilde{\bm{\phi}}$. From the
invariance of the generating functional ${\mathcal Z}[j]$ with respect
to a change of variables
$\bm{\phi}\to\bm{\phi}+\delta\bm{\phi}$ and
$\tilde{\bm{\phi}}\to\tilde{\bm{\phi}}+\delta\tilde{\bm{\phi}}$
with arbitrary (smooth)  functions $\delta\bm{\phi}$ and
$\delta\tilde{\bm{\phi}}$ we may then conclude that
\begin{equation}
  \label{eq:eqmotion}
  \left\langle\delta{\mathcal J}+\delta{\mathcal J}_j\right\rangle_{j}=0\;.
\end{equation}
Here $\delta{\mathcal J}$ and $\delta{\mathcal J}_j$ are the implied
changes of first order in $\delta\bm{\phi}$ and
$\delta\tilde{\bm{\phi}}$ of the action ${\mathcal J}$ and its
source part ${\mathcal J}_j$, respectively, and $\langle.\rangle_j$
denotes the average in the presence of sources.

Explicitly, we have
\begin{eqnarray}
  \label{eq:eqmocur}
  \delta{\mathcal J}&=&\int_{-\infty}^{\infty}{\Biggl\{
      \int_{{\mathbb{R}}^d_+}{\left[
          {\mathcal J}_{\bm{\phi}}\,\delta\bm{\phi}+
          {\mathcal J}_{\tilde{\bm{\phi}}}\,\delta\tilde{\bm{\phi}}\right]}}\nonumber\\
&&\mbox{}+\int_{\mathcal B}{\Big[
      {\mathcal J}_{{\bm{\phi}}_{{\mathcal B}}}\,\delta\bm{\phi}+
          {\mathcal
            J}_{{\tilde{\bm{\phi}}}_{{\mathcal B}}}\,\delta\tilde{\bm{\phi}}}\nonumber\\
&&\mbox{}+{\mathcal
  J}_{{\partial_n}{\bm{\phi}}}\,{\partial_n}\delta\bm{\phi}+{\mathcal J}_{{\partial_n}\tilde{\bm{\phi}}}\,{\partial_n}\delta\tilde{\bm{\phi}}\nonumber\\&& \mbox{}+
{\mathcal
  J}_{(\triangle{\bm{\phi}})_{{\mathcal B}}}\,{\triangle}\delta\bm{\phi}+{\mathcal J}_{({\triangle}\tilde{\bm{\phi}})_{{\mathcal B}}}\,{\triangle}\delta\tilde{\bm{\phi}}
    \Big]
      \Biggr\}
\end{eqnarray}
with
\begin{eqnarray}
  \label{eq:Jphi}
{\bm{\mathcal J}}_{\bm{\phi}}&=&-\partial_t\tilde{\bm{\phi}}-\lambda_0\Big[-\triangle +{\bm{U}}_2
\Big]\triangle\tilde{\bm{\phi}}\nonumber\\&&\mbox{}-
\lambda_0\,f_0{\left[\triangle{\left(\tilde{\bm{\phi}}\times\bm{\phi}\right)}+\tilde{\bm{\phi}}\times{\mathcal H}_{\bm{\phi}}\right]}
\;,
\end{eqnarray}
\begin{eqnarray}
  \label{eq:Jphitilde}
 {\bm {\mathcal J}}_{\tilde{\bm{\phi}}}&=&\partial_t\bm{\phi}
-\lambda_0\,\triangle{\left({\mathcal H}_{\bm{\phi}}-2\,\tilde{\bm{\phi}}
\right)}
-\lambda_0\,f_0\,{\mathcal H}_{\bm{\phi}}\times{\bm{\phi}}\;,
\end{eqnarray}
\begin{eqnarray}
  \label{eq:Jphis}
{\bm{\mathcal J}}_{{\bm{\phi}_{{\mathcal B}}}}&=&\big(c_0-\partial_n
\big){\tilde{\bm{\Phi}}}-
\lambda_0[{\bm{U}}_2+c_0\,\delta(0)]\partial_n\tilde{\bm{\phi}}\nonumber\\
&&+\lambda_0\,f_0\,\tilde{\bm{\phi}}\times 
(c_0-\partial_n){\bm{\phi}}\;,
\end{eqnarray}
\begin{eqnarray}
  \label{eq:Jphistilde}
{\bm{\mathcal J}}_{\tilde{\bm{\phi}_{{\mathcal B}}}}&=&\lambda_0\,\partial_n
{\left(2\tilde{\bm{\phi}}-{\mathcal H}_{\bm{\phi}}\right)}
+\lambda_0f_0\,\bm{\phi}\times\partial_n\bm{\phi}\;,
\end{eqnarray}
\begin{equation}
  \label{eq:Jpartialnphi}
{\bm{\mathcal J}}_{\partial_n{\bm{\phi}}}=
\lambda_0\,\delta(0)\,\partial_n\tilde{\bm{\phi}}\;,
\end{equation}
\begin{equation}
  \label{eq:Jpartialnphitilde}
{\bm{\mathcal J}}_{\partial_n\tilde{\bm{\phi}}}=
\lambda_0\,\delta(0)\,(\partial_n-c_0){\bm{\phi}}\;,
\end{equation}
\begin{equation}
  \label{eq:JDeltaphis}
{\bm{\mathcal J}}_{(\triangle{\bm{\phi}})_{{\mathcal B}}}=
\lambda_0\,\partial_n\tilde{\bm{\phi}}\;
\end{equation}
and
\begin{equation}
  \label{eq:JDeltaphistilde}
{\bm{\mathcal J}}_{(\triangle\tilde{\bm{\phi}})_{{\mathcal B}}}=
\lambda_0\,(\partial_n-c_0){\bm{\phi}}\;.
\end{equation}
The subscript ${\mathcal{B}}$ at local quantities indicates
their restriction to the boundary. Further,
${\bm{U}}_2$ is the matrix of second derivatives of the $|\bm{\phi}|^4$
potential:
\begin{equation}
  \label{eq:u2}
  {U}_2^{\alpha\beta}=\tau_0\,\delta^{\alpha\beta}+\frac{u_0}{6}\,{\left(\delta^{\alpha\beta}+2\,\phi^\alpha\phi^\beta\right)}\;,
\end{equation}
and for convenience, we have introduced the field
\begin{equation}
  \label{eq:defPhitilde}
  \tilde{\bm{\Phi}}\equiv
{{\bm{\mathcal R}}^{\text{T}}}\cdot\tilde{\bm{\phi}}=
{\lambda_0}{\left(-\triangle\tilde{\bm{\phi}}+f_0\,\tilde{\bm{\phi}}\times{\bm{\phi}}\right)}\;,
\end{equation}
where ${\bm{\mathcal R}}^{\text{T}}$ denotes the transposed reaction
operator, i.e., $({\mathcal R}^{\text{T}})^{\alpha\beta}={\mathcal
  R}^{\beta\alpha}$.  We shall refer to $\tilde{\bm{\Phi}}$ as
\emph{response field} since this is its known physical
significance\cite{Jan76,BJW76} (as can be read off again from our
Eqs.~(\ref{eq:Jh}) and (\ref{eq:delhRphitilde}) below); it should not
be confused with the auxiliary field $\tilde{\bm{\phi}}$.

For simplicity, let us include here in the source part of the
action merely bulk sources ${\bm{J}}({\bm{x}},t)$
and $\tilde{\bm{J}}({\bm{x}},t)$ as well as surface
sources ${\bm{J}}_1({\bm{x}}_\|,t)$ and
${\tilde{\bm{J}}}_1({\bm{x}}_\|,t)$ which couple to
${\bm{\phi}}$, $\tilde{\bm{\phi}}$,
${\bm{\phi}}_{{\mathcal B}}$, and $\tilde{\bm{\phi}}_{{\mathcal B}}$,
respectively. Owing to the arbitrariness of
$\delta{\bm{\phi}}$ and $\delta\tilde{\bm{\phi}}$, it
follows from the above results that the `equations of motion'
\begin{equation}
  \label{eq:eqmJphi}
  {\bm{\mathcal
      J}}_{\bm{\phi}}({\bm{x}},t)={\bm{J}}({\bm{x}},t)\;,\quad{\bm{x}}\notin{\mathcal B}\;,
\end{equation}
\begin{equation}
  \label{eq:eqmJphitilde}
  {\bm{\mathcal
      J}}_{\tilde{\bm{\phi}}}({\bm{x}},t)=\tilde{\bm{J}}({\bm{x}},t)\;,\quad{\bm{x}}\notin{\mathcal B}\;,
\end{equation}
and the `boundary equations of motion'
\begin{eqnarray}
  \label{eq:boundeqm}
&&{\bm{\mathcal
      J}}_{\rho}({\bm{x}}_\|f,t)={\bm{J}}_1({\bm{x}}_{\|},t)\,\delta_{\rho,{\bm{\phi}}_{{\mathcal B}}}+
\tilde{\bm{J}}_1({\bm{x}}_\|,t)\,\delta_{\rho,\tilde{\bm{\phi}}_{{\mathcal B}}}\;,\nonumber\\[-0.5em]&&\\[-0.5em]&&
\rho ={\bm{\phi}}_{{\mathcal B}},\,\tilde{\bm{\phi}}_{{\mathcal B}},\,
\partial_n{\bm{\phi}},\,\partial_n\tilde{\bm{\phi}},\,
(\triangle{\bm{\phi}})_{{\mathcal B}},\,(\triangle\tilde{\bm{\phi}})_{{\mathcal B}}\;,\nonumber
\end{eqnarray}
hold inside of averages with the action ${\mathcal
  J}+{\mathcal{J}}_j$.  From the latter and
Eqs.~(\ref{eq:Jphis})--(\ref{eq:JDeltaphistilde}), we get the
previously given boundary conditions for ${\bm{\phi}}$,
Eqs.~(\ref{eq:stbcphi}) and (\ref{eq:scc}), and two additional ones for
$\tilde{\bm{\phi}}$:\, namely
\begin{equation}
  \label{eq:bcphitilde}
  \partial_n\tilde{\bm{\phi}}=0
\end{equation}
and
\begin{eqnarray}
  \label{eq:bcrespfield}
  ({\partial_n}-c_0)\tilde{\bm{\Phi}}
&=&\nonumber\\\mbox{}({\partial_n}-c_0)\,{\lambda_0}{\left(-\triangle\tilde{\bm{\phi}}+f_0\,\tilde{\bm{\phi}}\times{\bm{\phi}}\right)}
&=&0\;.
\end{eqnarray}

\subsubsection{Fluctuation-dissipation theorem}
\label{sec:fdt}

The significance of the boundary conditions (\ref{eq:bcphitilde}) and
(\ref{eq:bcrespfield}) can be understood as follows. Let us add a
(possibly time-dependent) magnetic field term to the Hamiltonian
(\ref{eq:HGL}), making the replacement
\begin{equation}
  \label{eq:HGLh}
  {\mathcal H}\to {\mathcal H}_{\bm{h}}={\mathcal H}-\int_{{\mathbb{R}}^d_+}{\bm{h}}\cdot{\bm{\phi}}\;,
\end{equation}
where we assume $\bm{h}({\bm{x}},t)$ to have support
only off the surface. This induces the change
\begin{equation}
  \label{eq:Jh}
 {\mathcal J}\to {\mathcal J}_{{\bm{h}}}= {\mathcal J}
-\int_{{\mathbb{R}}^d_+}{\bm{h}}\cdot{\tilde{\bm{\Phi}}}
\end{equation}
of the dynamic action. Hence we recover the usual correspondence
(known from the bulk case\cite{BJW76})
\begin{equation}
  \label{eq:delhRphitilde}
  \frac{\delta}{\delta{\bm{h}}({\bm{x}},t)}\leftrightarrow
\tilde{\bm{\Phi}}({\bm{x}},t)
\end{equation}
between functional derivatives with respect to
${\bm{h}}({\bm{x}})$, taken at
$\bm{h}=\bm{0}$, and insertions of the response
operator on the right-hand side.

Furthermore, the fluctuation-dissipation relation
\begin{equation}
  \label{eq:FDT}
 -\theta(t-t')\,{\big\langle\dot{\phi}^\alpha({\bm{x}},t)\,{\bm{\phi}}^\beta({\bm{x}}',t')\big\rangle}%\nonumber\\
=\big\langle{\phi}^\alpha({\bm{x}},t)\,
\tilde{\Phi}^\beta({\bm{x}}',t')\big\rangle
\end{equation}
can be derived as in Refs.~\onlinecite{Jan76}, \onlinecite{BJW76},
\onlinecite{Jan79}, and \onlinecite{Jan92}, owing to the form
(\ref{eq:Jdbf}) of the action.

The significance of the boundary condition (\ref{eq:bcrespfield})
becomes clear if we let ${\bm{x}'}$ in Eq.~(\ref{eq:FDT})
approach a point on the boundary plane ${{\mathcal B}}$: it ensures
the consistency of the fluctuation-dissipation theorem (\ref{eq:FDT})
with the boundary condition (\ref{eq:stbcphi}).

To understand the Neumann boundary condition (\ref{eq:bcphitilde}),
note first that according to Eq.~(\ref{eq:Jdbf}), the reaction
operator couples $\nabla\tilde{\bm{\phi}}$ to the current
operator $-\lambda_0\,\nabla\delta{\mathcal
  H}/\delta{\bm{\phi}}$. This boundary condition ensures that
the reaction operator is self-adjoint. In addition, it can be combined
with Eq.~(\ref{eq:Jphistilde}) to see that the boundary equation of
motion (\ref{eq:boundeqm}) for
$\rho=\tilde{\bm{\phi}}_{{\mathcal B}}$ leads back to the
boundary conditions (\ref{eq:scc}) for the currents.

\section{RG analysis of the semi-finite model J}
\label{sec:RG}

\subsection{Preliminaries}
\label{sec:prelim}

We now turn to the RG analysis of the semi-infinite model J introduced
in the previous section. To this end, two RG schemes will be used: a
massless one based on minimal subtraction of poles and the expansion
about six dimensions, called RS$_1$, and a massive one for fixed
dimensions $2<d\le 4$, called RS$_2$.

Before embarking on a discussion of either one of these, we must set up
some notation. Let us define the generating functionals of (connected)
correlation and response functions
\begin{eqnarray}
  \label{eq:Wdef}
 \lefteqn{{\mathcal{W}}{\big[\tilde{\bm{J}},\bm{J},\bm{K};{\tilde{\bm{J}}}_1,\bm{J}_1\big]}}&&
\nonumber\\
&=&\ln{\left\langle e^{(\tilde{\bm{J}},\tilde{\bm{\phi}})+({\bm{J}},{\bm{\phi}})+(\bm{K},\tilde{\bm{\phi}}\times\bm{\phi})+({\tilde{\bm{J}}}_1,{\tilde{\bm{\phi}}_{\mathcal B}})+({\bm{J}}_1,{\bm{\phi}}_{\mathcal B})}\right\rangle}
\end{eqnarray}
and
\begin{eqnarray}
  \label{eq:Gdef}
 \lefteqn{{\mathcal{G}}{\big[\tilde{\bm{J}},\bm{J};{\tilde{\bm{J}}}_1,\bm{J}_1\big]}}&&
\nonumber\\
&=&\ln {\left\langle e^{(\tilde{\bm{J}},\tilde{\bm{\Phi}})+({\bm{J}},{\bm{\phi}})
+({\tilde{\bm{J}}}_1,{\tilde{\bm{\Phi}}_{\mathcal B}}) 
+({\bm{J}}_1,{\bm{\phi}}_{\mathcal B})}\right\rangle}\;,
\end{eqnarray}
where we have introduced the convenient short-hands
\begin{equation}
  \label{eq:Jphinot}
  ({\bm{J}},{\bm{\phi}})\equiv{\int_{-\infty}^{\infty}}dt\,
\int_{{\mathbb{R}}^d_+}d^dx\,{\bm{J}}({\bm{x}},t)\cdot{\bm{\phi}}({\bm{x}},t)
\end{equation}
and
\begin{equation}
  \label{eq:JphiBnot}
  ({\bm{J}}_1,{\bm{\phi}}_{\mathcal{B}})\equiv {\int_{-\infty}^{\infty}}dt\,
\int_{{\mathcal{B}}}d^{d-1}x_{\|}\,{\bm{J}}_1({\bm{x}}_\|,t)\cdot{\bm{\phi}}_{\mathcal{B}}({\bm{x}}_\|,t)\;.
\end{equation}
For the cumulants generated by these functionals we write
\begin{eqnarray}
  \label{eq:Wfuncdef}
\lefteqn{W^{(\tilde{N},N,L;\tilde{M},M)}}&&\nonumber\\
&=&\biggl\langle \prod_{j=1}^{\tilde{N}}
\tilde{\phi}^{\tilde{\alpha}_j}
\prod_{k=1}^N \phi^{\alpha_k}
\prod_{l=1}^{L}
(\tilde{\bm{\phi}}\times\bm{\phi})^{\gamma_l}
\prod_{m=1}^{\tilde{M}}
\tilde{\phi}_{\mathcal{B}}^{\tilde{\beta}_m}
\prod_{n=1}^M \phi_{\mathcal{B}}^{\beta_n}
\biggr\rangle^{\text{cum}}\nonumber\\
\end{eqnarray}
and
\begin{eqnarray}
  \label{eq:Gfuncdef}
  \lefteqn{ G^{(\tilde{N},N;\tilde{M},M)}}&&\nonumber\\
&=&\biggl\langle \prod_{j=1}^{\tilde{N}}
\tilde{\Phi}^{\tilde{\alpha}_j}
\prod_{k=1}^N \phi^{\alpha_k}
\prod_{m=1}^{\tilde{M}}
\tilde{\Phi}_{\mathcal{B}}^{\tilde{\beta}_m}
\prod_{n=1}^M \phi_{\mathcal{B}}^{\beta_n}
\biggr\rangle^{\text{cum}}\;.
\end{eqnarray}
We normally suppress the tensorial indices
$\tilde{\alpha}_1,\ldots,\beta_M$ of these functions, but give their
space and time (or momentum and frequency) coordinates (suppressed
above) when dealing with specific ones.

The reason for considering the functions
$W^{(\tilde{N},N,L;\tilde{M},M)}$ should be clear: aside from
multi-point cumulants of the basic fields $\tilde{\bm{\phi}}$ and
$\bm{\phi}$, insertions of the composite operator
$\tilde{\bm{\phi}}\times\bm{\phi}$ are needed because it appears in
the fluctuation-dissipation relation (\ref{eq:FDT}).

The free response propagator and free correlator one needs to compute
the $W$ and $G$ functions defined above are the same as for model
B$_{\text{B}}$, and for the case $\tau_0>0$, may be gleaned from any
of the Refs.~\onlinecite{DJ92,Die94b,WD95}. In a mixed
${\bm{p}}z\omega$ representation (where
${\bm{p}}\in{\mathbb{R}}^{d-1}$ is a ($d-1$)-dimensional wave-vector
conjugate to ${\bm{x}}_\|$ while $\omega$ denotes the Fourier
frequency variable associated with $t$), the free response propagator
reads
\begin{widetext}
\begin{eqnarray}
  \label{eq:freeG}
  \hat{G}(\bm{p};z,\tilde{z};\omega)&=&\frac{1}{2\lambda_0}\,
{\Big(\frac{\tau_0}{4}+i\frac{\omega}{\lambda_0}\Big)}^{-1/2}\,{\Big\{}
\frac{1}{2\kappa_-}\,{\big[}e^{-\kappa_-|z-\tilde{z}|}
-f_-\,e^{-\kappa_-(z+\tilde{z})}-g_-\,e^{-(\kappa_-z
+\kappa_+\tilde{z})}{\big]}\nonumber\\
&&\qquad\mbox{}-\frac{1}{2\kappa_+}\,{\big[}
e^{-\kappa_+|z-\tilde{z}|}-f_+\,e^{-\kappa_+(z+\tilde{z})}
-g_+\,e^{-(\kappa_+z+\kappa_-\tilde{z})}
{\big]}{\Big\}}
\end{eqnarray}
with
\begin{equation} \label{eq:fpm}
f_\pm \equiv f_\pm(\kappa_\pm,\kappa_\mp;c_0,\kappa) = 
\frac{{\kappa_\pm}\,{\kappa_\mp}\,(\kappa_\pm^2-\kappa_\mp^2)
- c_0\,{[{\kappa_\pm}\,{(\kappa^2-\kappa_\pm^2)}
+ {\kappa_\mp}\,{(\kappa^2-\kappa_\mp^2)}]}}
{{\kappa_\pm}\,{(\kappa^2-\kappa_\pm^2)}{(c_0+\kappa_\mp)}-{\kappa_\mp}
{(\kappa^2-\kappa_\mp^2)}{(c_0+\kappa_\pm)}}
\end{equation}
and
\begin{equation} \label{eq:gpm}
g_\pm\equiv g_\pm(\kappa_\pm,\kappa_\mp;c_0,\kappa)
=\frac{2c_0\,{\kappa_\pm}\,{(\kappa^2-\kappa_\mp^2)}}
{{\kappa_\pm}\,{(\kappa^2-\kappa_\pm^2)}
{(c_0+\kappa_\mp)}-{\kappa_\mp}\,{(\kappa^2-\kappa_\mp^2)}{(c_0+\kappa_\pm)}}
\;.
\end{equation}
\end{widetext}
Here
\begin{equation} \label{eq:kappa}
\kappa \equiv +\sqrt{{\bm{p}}^2+\tau_0}\;,
\end{equation}
and $\kappa_\pm$ denote the roots with positive real parts of the equation
\begin{equation}
  \label{eq:kappapm}
  \kappa_\pm^2={\bm{p}}^2+({\tau_0}/{2})\pm {\big[({\tau_0}/{2})^2
+i\,{(\omega/\lambda_0)}\big]}^{1/2}\;.
\end{equation}
The free correlator
$C\,\delta^{\alpha\beta}=\langle\phi^\alpha\phi^\beta\rangle^{\text{cum}}$ can be
expressed in terms of the free response propagator as
\begin{eqnarray}
  \label{eq:freeC}
  \hat{C}({\bm{p}};z,z';\omega)&=&2\lambda_0\int_0^\infty\! 
d\tilde{z}\,\hat{G}({\bm{p}};z,\tilde{z};\omega)\nonumber\\
&&\qquad\quad\times({\bm{p}}^2-\partial_{\tilde{z}}^2)\,
\hat{G}(-{\bm{p}};z',\tilde{z};-\omega)\;.\nonumber\\
\end{eqnarray}

Owing to the presumed $O(3)$ symmetry of the Hamiltonians
(\ref{eq:Hamil}) and (\ref{eq:HGL}) of our lattice and continuum
models, the only surface transition that is possible in three
dimensions is the \emph{ordinary} transition. We can benefit from the
fact that its asymptotic critical behavior can be studied by taking
the limit $c_0\to\infty$ (see, e.g., Ref.~\onlinecite{Diehl} and
below). The simplified expressions for the free response propagator
and correlator which then apply correspond to the replacements of the
coefficients $f_\pm$ and $g_\pm$ by
\begin{equation}
  \label{eq:fpminf}
 f^\infty_\pm\equiv \lim_{c_0\to\infty}f_\pm(\kappa_\pm,\kappa_\mp;
c_0,\kappa)=-\frac{\kappa_\pm-\kappa_\mp}{\kappa_\pm+\kappa_\mp}
\end{equation}
and
\begin{equation}
  \label{eq:gpminf}
 g^\infty_\pm\equiv \lim_{c_0\to\infty}g_\pm(\kappa_\pm,\kappa_\mp;
c_0,\kappa)=-\frac{2\kappa_\pm}{\kappa_\pm+\kappa_\mp}\;,
\end{equation}
respectively.

\subsection{Massless renormalization scheme near six dimensions (RS$_1$)}
\label{sec:RS1}

We here restrict ourselves to bulk dimensions $4<d\le 6$. Then the
static critical behavior is described by Landau theory. The Gaussian
fixed point, $u_0=0$, of the $|\bm{\phi}|^4$ theory is
infrared-stable. In part of the calculation $u_0$ therefore can be set
to zero. This is possible as long as we consider quantities that have
a nonsingular and nonvanishing limit $u_0\to 0$. However, we must keep
in mind that the linear scaling field $u$ associated with $u_0$ is
dangerous irrelevant.\cite{Fis83,AP82} Quantities like the free energy
density or the spontaneous magnetization vary as inverse powers of
$u_0\sim u$ for $u_0\to 0$, and hyperscaling is broken.  Accordingly,
already a full scaling description of the \emph{static bulk} critical
behavior requires the inclusion of a second, so-called thermodynamic,
length besides the bulk correlation length. Finally, in applying
techniques of renormalized field theory, we must remember that both
the static as well as the dynamic theories are not renormalizable for
$d>4$ if $u_0\ne 0$. Single insertions of the local operator to which
$u_0$ couples can be renormalized, but the additional counterterms to
which it gives rise are not sufficient for curing the additional
ultraviolet (uv) singularities produced by multiple insertions.

We now consider the $W$ functions (\ref{eq:Gfuncdef}), where we
restrict the temperature $T$ to values above the critical temperature
$T_{\text{c}}$ and set $u_0=0$ temporarily. Since the Hamiltonian
(\ref{eq:HGL}) becomes Gaussian for $u_0=0$, there are no static uv
singularities to cure. Hence no amplitude renormalization of the order
parameter is required, i.e., $Z_\phi=1$. By power counting one finds
that counterterms of the form ${\int}dt\,\int_{{\mathbb{R}}^d_+}
\tilde{\bm{\phi}}\cdot\dot{\bm{\phi}}$ are also not needed. Since this
implies that the product of amplitude renormalization factors
$Z_\phi\,Z_{\tilde{\phi}}$ for $\bm{\phi}$ and $\tilde{\bm{\phi}}$ is
unity, an amplitude renormalization of $\tilde{\bm{\phi}}$ is not
required either ($Z_{\tilde{\phi}}=1$). A (bulk) counterterm $\propto
\bm{\phi}\cdot\triangle\tilde{\bm{\phi}}$ is ruled out for $u_0=0$. If
$u_0$ did not vanish, the $O(u_0)$ contribution to such a counterterm
would diverge as $\Lambda^4$ for $d = 6$ within a theory regularized
by a cutoff $\Lambda$.

We assert that the following counterterms are sufficient to
renormalize the generating functional ${\mathcal W}$: aside from those
implied by the reparametrizations
\begin{equation}
  \label{eq:Zlamdefg4}
  \lambda_0=\mu^{-4}Z_\lambda(f^2,d)\,\lambda
\end{equation}
and
\begin{equation}
  \label{eq:Zfdefdg4}
  f_0=\mu^{\frac{6-d}{2}}\, {\big[Z_\lambda(f^2,d)\big]}^{-1}\,f\;,
\end{equation}
only a counterterm of the form
${\int}dt{\int_{{\mathbb{R}}^d_+}}\bm{K}\cdot\triangle\tilde{\bm{\phi}}$
is required, where $\mu$ is an arbitrary momentum scale. More
precisely, we claim that the cumulants generated by the functional
\begin{widetext}
\begin{equation}  
  \label{eq:Wren}
  {\mathcal{W}}_{\text{ren}}{\big[\tilde{\bm{J}},\bm{J},\bm{K};
{\tilde{\bm{J}}}_1,\bm{J}_1\big]}=
{\mathcal{W}}{\big[\tilde{\bm{J}},\bm{J},\bm{K};
{\tilde{\bm{J}}}_1,\bm{J}_1\big]}
-{\big(Z_\lambda-1\big)}f^{-1}\,\mu^{-\frac{6-d}{2}}
\int_{-\infty}^{\infty}\!dt\int_{{\mathbb{R}}^d_+}
\bm{K}\cdot\triangle\tilde{\bm{\phi}}
\end{equation}
\end{widetext}
are \emph{uv finite} when expressed in terms of $\lambda$, $f$ (and $c_0$ or
its dimensionless equivalent $c\equiv c_0/\mu$).

This conclusion is based on the following observations. The
detailed-balance form of the action (\ref{eq:Jdbf}) in conjunction
with the constraints imposed by the conservation of the order
parameter and power counting restricts the possible bulk counterterms
to those included in Eq.~(\ref{eq:Wren}). Using this result as input,
one can consider the renormalization of the bulk analog of the
fluctuation-dissipation relation (\ref{eq:FDT}).  For convenience we
employ dimensional regularization and fix the counterterms by minimal
subtraction of poles at $d=6$. From the uv finiteness of the
correlation function of the renormalized function on the left-hand
side of this equation two conclusions may then be drawn: the
renormalization factors of $f_0$ and $\lambda_0$ are reciprocal to
each other, and the renormalization function of the
${\bm{K}}$-dependent counterterm is related to $Z_\lambda$ in the
stated fashion.\cite{BJWcomins} The result means that insertions of
the response field (\ref{eq:defPhitilde}) renormalize just as
$\bm{\phi}$, requiring no additional counterterms.

A final step remains to complete the argument: we must show that no
further surface counterterms are needed. Given the causal structure of
the theory (according to which at least one $\tilde{\bm{\phi}}$ must
occur in any monomial of the action), power counting restricts the
possible candidates for such counterterms in $d=6$ dimensions to
boundary contributions to the action involving monomials of the form
$\lambda_0\,\tilde{\bm{\phi}}^2$,
$\lambda_0\,\tilde{\bm{\phi}}\cdot\bm{\phi}$, and
$\lambda_0\,{\tilde{\phi}}^\alpha\phi^\beta\phi^\gamma$ (as well as
similar ones with derivatives), where the coefficients have momentum
dimensions $1$, $3$, and $0$, respectively. Now the cubic nonlinearity of
the bare action can be rewritten as
\begin{equation}
  \label{eq:Jmc}
  {\mathcal{J}}_{\text{mcv}}=\lambda_0\,f_0\,\epsilon^{\alpha\beta\gamma}
{\int_{-\infty}^\infty}dt\int_{{\mathbb{R}}^d_+}
{\big(\bm{\nabla}\tilde{\phi}^\alpha\big)}\,
\phi^\beta\,\bm{\nabla}\phi^\gamma
\end{equation}
upon integrating by parts and utilizing the boundary condition
(\ref{eq:stbcphi}). Thus each $\tilde{\phi}$ leg of the vertex
$\propto f_0$ comes with a derivative $\bm{\nabla}$. This reduces the
superficial degree of divergence of such uv boundary singularities
with two $\tilde{\phi}$ legs by two, making it negative (uv superficially
convergent). By a similar argument, surface counterterms involving one
$\tilde{\phi}$  and two $\phi$ fields are ruled out. Hence we are left
with surface counterterms
$\propto\bm{\phi}\cdot\bm{\nabla}\tilde{\bm{\phi}}$ and analogous ones
with up to two additional derivatives.

To proceed we follow Ref.~\onlinecite{BJW76} and perform the
integral $\int_{-\infty}^\infty dt$ of the fluctuation-dissipation
relation (\ref{eq:FDT}). This yields
\begin{widetext}
\begin{eqnarray}
  \label{eq:FDTcons1}
  \big\langle\phi^\alpha(\bm{x})\,\phi^\beta(\bm{x}')\big\rangle^{\text{st}}&=&
-\lambda_0\,\triangle' \big\langle\phi^\alpha(\bm{x})
\tilde{\phi}^\beta(\bm{x}')\big\rangle_{\omega=0}+\lambda_0f_0\,\big\langle
\phi^\alpha(\bm{x})\,\big(\tilde{\bm{\phi}}\times\bm{\phi}\big)^\beta(\bm{x}')
\big\rangle_{\omega=0}\;,
\end{eqnarray}
where the superscript `st' indicates a static quantity while the
subscript $\omega$ at the expectation values on the right-hand side
means their Fourier transform with respect to the time difference
$t-t'$. We multiply this equation from the right by the inverse of the
static propagator on the right-hand side and from the left with the
vertex function $\Gamma_{\tilde{\phi}^\alpha\phi^\beta}$.
The result is\cite{comVF} 
\begin{equation}
  \label{eq:FDTvert}
  \Gamma_{\tilde{\phi}^\alpha\phi^\beta}(\bm{x},\bm{x}';\omega{=}0)=
\lambda_0\,(-\triangle'+\tau_0){\big[}-\triangle\,\delta(\bm{x}-\bm{x}')
+ f_0\,\Gamma_{\tilde{\phi}^\alpha;
\,(\tilde{\bm{\phi}}\times\bm{\phi})^\beta}(\bm{x},\bm{x}';\omega{=}0)\big],
\end{equation}
\end{widetext}
where
$\Gamma_{\tilde{\phi}^\alpha;\,(\tilde{\bm{\phi}}\times\bm{\phi})^\beta}$
means a vertex functions with a single $\tilde{\phi}^\alpha$ leg and
an insertion of the composite operator
$(\tilde{\bm{\phi}}\times\bm{\phi})^\beta$. Owing to the operator
$-\triangle'+\tau_0$ (inverse static propagator) on the right-hand
side, the uv behavior of possible primitive local divergences of the
vertex function on the left-hand side is improved by two powers of
$\Lambda$. This is sufficient to ensure that no additional surface
counterterms with one $\tilde{\phi}$ and one $\phi$ leg are needed.
Since the vertex function $\Gamma_{\tilde{\phi}^\alpha;
  \,(\tilde{\bm{\phi}}\times\bm{\phi})^\beta}$ has an explicit
$\bm{\nabla}$ acting on the external leg, it has a primitive
logarithmic bulk singularity that is cured by the subtraction provided
by the bulk counterterm $\propto \bm{K}\cdot
\triangle\tilde{\bm{\phi}}$.  On dimensional grounds one might
anticipate logarithmically divergent surface counterterms of the form
$ (\partial_n\tilde{\bm{\phi}})\cdot(\tau_0-\triangle)\bm{\phi}$ and
$\tilde{\bm{\phi}}\cdot\partial_n(\tau_0-\triangle)\bm{\phi}$, but
these are annihilated by the boundary conditions (\ref{eq:bcphitilde})
and (\ref{eq:scc}), respectively. Note also that the restriction to
$\omega=0$ is unproblematic here because each factor of $\omega$
(i.e., each time derivative) reduces the superficial degree of
divergence by four.

A further comment is appropriate here. Field theories for systems with
boundaries are known to have the following feature: Besides
one-particle \emph{irreducible} (1PI) graphs, one-particle
\emph{reducible} (1PR) ones may also require `final subtractions' and
hence contribute to renormalization functions (cf.\ Sec.~II.B.6 of
Ref.\ \onlinecite{Diehl}).  Nevertheless our above reasoning based on
1PI graphs is conclusive since the power counting would not be changed
if we contracted 1PR graphs whose external free legs are amputated to
a point. Hence the counterterms included in ${\mathcal
  W}_{\text{ren}}$ are indeed sufficient to cure the uv singularities
of the $W$ functions (\ref{eq:Wfuncdef}). The uv finiteness of the $G$
functions generated by
\begin{equation}
  \label{eq:Gren}
   {\mathcal{G}}_{\text{ren}}{\big[\tilde{\bm{J}},\bm{J},\bm{K};
{\tilde{\bm{J}}}_1,\bm{J}_1\big]}=
{\mathcal{G}}{\big[\tilde{\bm{J}},\bm{J},\bm{K};
{\tilde{\bm{J}}}_1,\bm{J}_1\big]}
\end{equation}
when expressed in terms of $\lambda$ and $f$ (as well as $\tau_0$ and
$c_0$ or their dimensionless equivalents $\tau\equiv\tau_0/\mu^2$ and
$c\equiv c_0/\mu$) follows as a simple corollary from the fact that
the response field $\tilde{\bm{\Phi}}$ renormalizes just as
$\bm{\phi}$ (namely trivially).

\subsection{RG analysis in $\bm{6-\epsilon}$ dimensions}
\label{sec:RGAeps6}

Utilizing the results of the previous subsection, one can perform a RG
analysis of quantities that are finite and nonzero for $u_0 = 0$.
This criterion is satisfied by both the $W$ and the $G$ functions
for $\tau_0>0$, as can be checked via perturbation theory in
$f_0$. Since such a RG analysis is standard we can be brief and just
state its principal results.

To one-loop order the renormalization function $Z_\lambda$ is given by
\cite{HKJlec}
\begin{equation}
  \label{eq:Zlambda}
  Z_\lambda(f)=1-
\frac{1}{192\,\pi^3}\,\frac{f^2}{6-d}
+O(f^4)\;.
\end{equation}
Equation (\ref{eq:Zfdefdg4}) implies that the beta function
\begin{equation}
  \label{eq:betafdef}
  \beta_f(f)\equiv\left.\mu\partial_\mu\right|_0f
\end{equation}
can be written in terms of the exponent function
\begin{equation}
  \label{eq:etalambdadef}
  \eta_\lambda(f)\equiv\left.\mu\partial_\mu\right|_0\ln Z_\lambda\;,
\end{equation}
namely
\begin{equation}
  \label{eq:betaf}
  \beta_f(f)=-f\,{\left[\frac{6-d}{2}-\eta_\lambda(f)\right]}\;,
\end{equation}
where $\left.\partial_\mu\right|_0$ denotes the derivative at fixed bare
coupling constants $f_0$, $\tau_0$, and $c_0$. The infrared-stable fixed
point $f^*$ for $4 \leq d < 6$ is given by the nontrivial root of the equation
$\beta_f(f^*) = 0$. From Eq.~(\ref{eq:betaf}) we find for the value of the
exponent function $\eta_\lambda^*$ at the fixed point $f^*$ the result
\begin{equation}
  \label{eq:etalambda*}
  \eta_\lambda^*\equiv \eta_\lambda(f^*)=\frac{6-d}{2}\;,
\end{equation}
which we insert into the general expression
\begin{equation}
  \label{eq:dynzgdef}
  \mathfrak{z}=4-\eta_\lambda^*
\end{equation}
for the dynamic critical exponent $\mathfrak{z}$. Since the
correlation exponent $\eta$ is zero for $d\ge 4$, the final result
$\mathfrak{z}=(d+2)/2$ for $4\le d\le 6$ is consistent with the
established value\cite{Wag70,MM75,Kaw75,Jan76,BJW76,Doh76}
\begin{equation}
  \label{eq:dynz}
  \mathfrak{z}=\frac{d+2-\eta}{2}\;,\quad 2<d\le 6\;. 
\end{equation}
The latter result is known to follow most easily from the observation
that the characteristic frequency of isotropic ferromagnets for
$T>T_{\text{c}}$ is the Larmor frequency and hence scales as the
(static) bulk magnetic field $\bm{h}$.\cite{Wag70,spwavefreq}

Next we turn to the analysis of the critical behavior of
\emph{surface} quantities. We restrict the following discussion to
the case of the \emph{ordinary} transition, the only surface transition
at bulk criticality that remains in the physically interesting
three-dimensional case.\cite{rem:normal,DE84} The surface enhancement
variable $c\equiv c_0/\mu$ transforms as $c\to \bar{c}(l)=c/\ell$
under scale transformations $\mu\to \mu\,\ell$ and hence approaches
the fixed-point value $c^*_{\text{ord}}=\infty$ in the infrared limit
$\ell\to 0$, provided its initial value is positive. In this case we can
set $c_0=\infty$ from the outset. Surface quantities involving the surface
fields $\phi_{\mathcal{B}}$ or $\tilde{\bm{\Phi}}_{\mathcal{B}}$ then
vanish.

Let $F_{\text{ren}}$ be any of the two renormalized functions
$W^{(\tilde{N},N,L;\tilde{M},M)}_{\text{ren}}$ and
$G_{\text{ren}}^{(\tilde{N},N;\tilde{M},M)}$ generated by the
functionals (\ref{eq:Wren}) and (\ref{eq:Gren}) for $u_0=0$ and
$\tau\equiv\tau_0/\mu^2\ge 0$. The invariance of the regularized bare
functions with respect to a change of $\mu$ implies the RG equation
\begin{equation} \label{eq:RGEeps}
\left[\mu\partial_\mu + \beta_f \partial_f - 2\tau \partial_\tau
+(4-\eta_\lambda)\lambda \partial_\lambda - c\partial_c\right]
F_{\text{ren}} = 0\;.
\end{equation}
This may be utilized in a familiar manner to obtain the asymptotic
scaling forms of these functions. The result one obtains for the pair
correlation function $C^{\alpha\beta}=\delta^{\alpha \beta}\, C$
($=G^{(0,2;0,0)}$) at $T=T_{\text{c}}$ agrees with the more general
one predicted in our previous paper,\cite{KKD01}
\begin{equation}\label{eq:Czzrt}
C(\bm{r};z,z';t)\approx
r^{2-d-\eta}\,\Upsilon{\left({z/ r},{z'/ r};
t{r}^{-{\mathfrak{z}}}\right)}\;,
\end{equation}
if the classical value $\eta=0$ and the implied one
$\mathfrak{z}=(d+2)/2$ are substituted (as is appropriate for $4<d<6$).
Here we have suppressed the variables $\mu$ and $\lambda$, setting
both to unity. The variable $c$ does not appear on the right-hand side
because the scaling function $\Upsilon$ is a property of the
$c=\infty$ fixed point. Deviations of $c$ from the value $c=\infty$
produce corrections to the displayed leading infrared contribution.

The derivation of the scaling form of the surface correlation function
\begin{equation}
  \label{eq:scf}
  C_{11}(\bm{r};t)\equiv C(\bm{r};0,0;t)
\end{equation}
is not quite so straightforward because $C|_{c=\infty}$ vanishes
whenever $z=0$ or $z'=0$, as a consequence of the Dirichlet boundary
condition into which Eq.~(\ref{eq:stbcphi}) turns for $c_0=\infty$.
One possibility to cope with this difficulty is to the study behavior
of $C_{11}$ in the limit $c\to \infty$. As is expounded
elsewhere,\cite{Diehl,DD80,DJ92,Die94b,WD95} this can be achieved by
an expansion in powers of $1/c_0$.  According to the boundary conditions
(\ref{eq:stbcphi}) and (\ref{eq:bcrespfield}), the boundary operators
$\bm{\phi}_{\mathcal{B}}$ and $\tilde{\bm{\Phi}}_{\mathcal{B}}$ can be
replaced by $c_0^{-1}\partial_n\bm{\phi}_{\mathcal{B}}$ and
$c_0^{-1}\partial_n\tilde{\bm{\Phi}}_{\mathcal{B}}$, respectively.
From previous detailed investigations\cite{Diehl,DD80,Die94b} of the
$1/c_0$ expansion it is therefore clear how the scaling forms that the
correlation and response functions involving these boundary operators
take at the ordinary transition can be determined: making the
replacements
$\bm{\phi}_{\mathcal{B}}\to\partial_n\bm{\phi}_{\mathcal{B}}$ and
$\tilde{\bm{\Phi}}_{\mathcal{B}}\to
\partial_n\tilde{\bm{\Phi}}_{\mathcal{B}}$, one studies the
so-obtained analogs of these functions, with $c_0$ set to $\infty$.

An alternative strategy, which leads to equivalent results, is to use
the boundary operator expansion (BOE). According to Eqs.~(\ref{eq:stbcphi})
and (\ref{eq:bcrespfield}), both the order parameter density
$\bm{\phi}$ as well as the response field $\tilde{\bm{\Phi}}$ satisfy
Dirichlet boundary conditions if $c_0=\infty$. In analogy with the
static case,\cite{DD81a,Diehl} a BOE of the form
\begin{equation}
  \label{eq:boephi}
  \bm{\phi}(\bm{x},t)\mathop{\approx}\limits_{z\to 0}
 D(z,t)\,\partial_n\bm{\phi}(\bm{x}_{\mathcal{B}},t)+ \ldots
\end{equation}
and its counterpart involving $\tilde{\bm{\Phi}}$ and
$\partial_n\tilde{\bm{\Phi}}$ should hold as $\bm{x}=(\bm{x}_\|,z)$
approaches the surface point $\bm{x}_{\mathcal{B}}=(\bm{x}_\|,0)$. (We
have suppressed contributions proportional to the operator $\openone$,
which occur when the average $\langle\bm{\phi}\rangle$ does not
vanish.\cite{Diehl,Sym81,KED95}) Equation \ref{eq:boephi} implies that
cumulants involving the surface operators $\partial_n\bm{\phi}$ and
$\partial_n\tilde{\bm{\Phi}}$ give access to surface correlation functions.
We refrain from doing this within the framework of the $\epsilon_6$
expansion and turn directly to RS$_2$, the massive RG scheme.

\subsection{Massive renormalization scheme for $\bm{2<d\le 4}$ (RS$_2$)}
\label{sec:RS2}

Our aim here is to extend the massive RG scheme for semi-infinite
systems developed by one of us and Shpot\cite{DS94,DS98} to the
dynamic theory of the model-J action (\ref{eq:JmodJ}). We
assume that $2<d\le 4$ and give up the restriction $u_0=0$, i.e.,
both $u_0$ and $f_0$ are assumed to be nonzero.

The scheme can be extended for general values of $c_0$.  However, as
we are primarily interested in the dynamic surface critical behavior
at the \emph{ordinary} transition in $d=3$ bulk dimensions, we can
simplify the analysis by setting $c_0=\infty$ from the outset. The
advantage of doing this is considerable: for general values of $c_0$,
the renormalization factors associated with surface operators (called
`surface renormalization factors' for short) depend on the
renormalized coupling constant $u$ (to be defined below) \emph{and}
the ratio $c/m$, where $c$ and $m$ are the renormalized
analog\cite{rem:rense} of the bare surface enhancement $c_0$ and the
renormalized mass $m$ (to be introduced below), respectively. This
additional dependence on the mass ratio $c/m$ makes the RG analysis
rather cumbersome. If we set $c_0=\infty$, we focus directly on the
asymptotic regime $c/m=\infty$ and avoid this difficulty because the
surface renormalization factors (which are of purely static origin)
can then be chosen to depend merely on $u$.

\subsubsection{Static bulk renormalization functions}\label{sec:stbrfs}

Let $\Gamma_{\text{st,b}}^{(N,I)}$ be the static bulk vertex function
with $N$ legs of type $\phi$ and $I$ insertions of $\bm{\phi}^2/2$, and
$\check{\Gamma}_{\text{st,b}}^{(N,I)}(\bm{q},\bm{Q})$ the Fourier 
transform of this (translationally invariant) function, up to the
momentum-conserving factor $(2\pi)^d\,\delta(\sum\bm{q}+\sum\bm{Q})$.
Here $\bm{q}$ and $\bm{Q}$ are the $N$ and $I$ momenta conjugate to
the positions $\bm{x}$ and $\bm{X}$ of the legs and the inserted
operators, respectively. We write
\begin{equation}
  \label{eq:mshift}
  \tau_0=m^2+\delta m^2\;,
\end{equation}
\begin{equation}
  \label{eq:phiren}
  \bm{\phi}=[Z_\phi(u)]^{1/2}\,\bm{\phi}_{\text{ren}}\;,
\end{equation}
\begin{equation}
  \label{eq:phi2ren}
  \bm{\phi}^2=[Z_{\phi^2}(u)]^{-1}\,\big[\bm{\phi}^2\big]_{\text{ren}}\;,
\end{equation}
and
\begin{equation}
  \label{eq:uren}
  u_0=Z_u(u)\,m^{4-d}\,u\;,
\end{equation}
which introduce the renormalized mass $m$, the renormalized densities
$\bm{\phi}_{\text{ren}}$ and $[\bm{\phi}^2]_{\text{ren}}$, and the
renormalized coupling constant $u$.

The mass shift $\delta m^2$ and the renormalization (`$Z$') factors
are fixed via the familiar normalization conditions
\begin{equation}\label{m2}
\left.\check\Gamma^{(2)}_{\text{st,b,ren}}\!\left(q;u,m\right)
\right|_{q
=0}=m^2\;,
\end{equation}
\begin{equation}\label{dgq}
\left.\frac{\partial}{\partial q^2}\,
\check\Gamma^{(2)}_{\text{st,b,ren}}\!\left(q;u,m\right)
\right|_{q=0}=1\;,
\end{equation}
\begin{equation}\label{ng21}
\left.\check
\Gamma^{(2,1)}_{\text{st,b,ren}}\!
\left(\bm{q},\bm{Q};u,m\right)\right|_{\bm{q}=\bm{Q}=\bm{0}}=1\;,
\end{equation}
and
\begin{equation}\label{normgamma4}
\left.\check
\Gamma^{(4)}_{\text{st,b,ren}}\!\left(\{\bm{q}_i\};u,m\right)
\right|_{\{\bm{q}_i=\bm{0}\}}=
m^{4-d}\, u\
\end{equation}
for the renormalized static bulk (b) vertex functions 
\begin{equation}
  \label{eq:renstbvfs}
  \Gamma^{(N,I)}_{\text{st,b,ren}}(\cdot;m,u)
=[Z_\phi(u)]^{{N}/{2}}\,[Z_{\phi^2}(u)]^I\,
\Gamma^{(N,I)}_{\text{st,b}}(\cdot;\tau_0,u_0)
\end{equation}
with $(N,I)\ne (0,1),\,(0,2)$.

\subsubsection{Dynamic bulk renormalization functions}\label{sec:dynbrfs}

We introduce renormalized auxiliary and response fields
$\tilde{\bm{\phi}}_{\text{ren}}$ and $\tilde{\bm{\Phi}}_{\text{ren}}$,
a dynamic bulk renormalization factor
 $Z_\lambda$, and the
renormalized bulk variables $\lambda$ and $f$ such that
\begin{equation}
  \label{eq:phitilderen}
  \tilde{\bm{\phi}}=[Z_\phi(u)]^{-1/2}\,\tilde{\bm{\phi}}_{\text{ren}}\;,
\end{equation}
\begin{equation}
  \label{eq:Phitilderen}
  \tilde{\bm{\Phi}}=[Z_\phi(u)]^{1/2}\,\tilde{\bm{\Phi}}_{\text{ren}}\;,
\end{equation}
\begin{equation}
  \label{eq:lambdaren}
  \lambda_0=m^{-4}\,Z_\lambda(u,f)\,\lambda\;,
\end{equation}
and
\begin{equation}
  \label{eq:fren}
  f_0=m^{\frac{6-d}{2}}\,[Z_\lambda(u,f)]^{-1}\,[Z_\phi(u)]^{1/2}\,f\;.
\end{equation}
To fix the function $Z_\lambda$, we choose the normalization condition
\begin{equation}
  \label{eq:normlambda}
  \left.\frac{\partial}{\partial q^2}\,
\check\Gamma_{\tilde{\phi}^\alpha\tilde{\phi}^\beta}^{\text{(b,ren)}}\!\left(q,\omega;u,f,\lambda,m\right)
\right|_{q=\omega=0}=-2\,m^{-4}\,\lambda\,\delta^{\alpha\beta},
\end{equation}
where
$\check\Gamma_{\tilde{\phi}^\alpha\tilde{\phi}^\beta}^{\text{(b,ren)}}$
denotes a renormalized dynamic bulk vertex function in the
momentum-frequency (${\bm{q}}\omega$) representation. 

Let us add a few clarifying remarks. Note, first, that
the renormalization functions introduced above are sufficient to
absorb the uv singularities of the vertex functions with arbitrary
numbers $N$ and $\tilde{N}$ of $\phi$ and $\tilde{\phi}$ legs of the
dynamic \emph{bulk} theory for $d\le 4$. Hence the above
reparametrizations also yield uv finite renormalized functions when
applied to the bulk analogs of the $\tilde{N}+N$ point cumulants with
$\tilde{N}$ $\tilde{\phi}$ and $N$ $\phi$ fields , i.e., to the bulk
analogs of the functions $W^{(\tilde{N},N,0;0,0)}$ defined in
Eq.~(\ref{eq:Wfuncdef}). The same remark applies to the bulk analogs
of the $G$ functions (\ref{eq:Gfuncdef}).

The meaning of the multiplicative renormalizability of the response
field $\tilde{\bm{\Phi}}$ with respect to the renormalization of the
$N+\tilde{N}$ point bulk vertex functions with an insertion of the
composite operator $\tilde{\bm{\phi}}\times\bm{\phi}$ has been
discussed elsewhere \cite{BJW76,Jan79} and need not be reiterated
here: it implies, in particular, that the renormalization of
$\Gamma^{\text{(b)}}_{\tilde{\phi}^\alpha;
  (\tilde{\bm{\phi}}\times\bm{\phi})^\beta}$ involves a subtraction
proportional to
$\Gamma^{\text{(b)}}_{\tilde{\phi}^\alpha;\triangle\phi^\beta}$,
chosen in conformity with the multiplicative renormalization
(\ref{eq:Phitilderen}).

Our second remark concerns the renormalization of
$\tilde{\bm{\phi}}_{\text{ren}}$ and $f$ in
Eqs.~(\ref{eq:phitilderen}) and (\ref{eq:fren}). The fact that no
primitive uv singularities involving a counterterm proportional to
$\tilde{\phi}^\alpha\dot{\phi}^\beta$ occur implies that the
renormalization factor of $\tilde{\bm{\phi}}_{\text{ren}}$ is given by
$Z_\phi^{-1/2}$ (up to a uv finite factor). This result means that the
latter product of operators transforms according to its engineering
dimension under RG transformations and is related to the conservation
of the order parameter.\cite{rem:opcons}

The form of the renormalization factor of $f$,
$Z_\lambda^{-1}\,Z_\phi^{1/2}$, follows from the
fluctuation-dissipation theorem (\ref{eq:FDT}); it tells us that
$\dot{\phi}^\alpha/\tilde{\Phi}^\alpha$ is a RG invariant. As a direct
consequence, a relation generalizing Eq.~(\ref{eq:betaf}) holds
between the beta function $\beta_f$ and exponent functions, which now
are defined via
\begin{equation}
  \label{eq:betafmassdef}
  \beta_f(u,f)\equiv\left.m\partial_m\right|_0f\;,\quad 
  \beta_u(u)\equiv\left.m\partial_m\right|_0u\;
\end{equation}
and
\begin{equation}
  \label{eq:etalambdamassdef}
  \eta_\kappa(u,f)\equiv\left.m\partial_m\right|_0\ln Z_\kappa\;,\quad
  \kappa=\lambda,\,\phi,\,\phi^2\,,
\end{equation}
respectively. We have
\begin{equation}
  \label{eq:betafmass}
  \beta_f(u,f)=-f\,{\left[\frac{6-d}{2}-\eta_\lambda(u,f)+
  \frac{\eta_\phi(u)}{2}\right]}\;.
\end{equation}
Since $\eta=\eta_\phi(u^*)$, where $u^*$ is the nontrivial zero of
$\beta_u$, this form ensures that the established result
(\ref{eq:dynz}) for the dynamic exponent $\mathfrak{z}$ is obtained if
the value 
\begin{equation}
  \label{eq:etalambdamassstar}
\eta_\lambda^*\equiv\eta_\lambda(u^*,f^*)=\frac{6-d+\eta}{2} 
\end{equation}
pertaining to the infrared-stable fixed point $(u^*,f^*)$ is
substituted into Eq.~(\ref{eq:dynzgdef}).

Finally, let us note that the renormalization factors $Z_\phi$,
$Z_{\phi^2}$, $Z_u$, $Z_{\lambda}$ introduced above are all \emph{uv
  finite} for $d<4$, although they are logarithmically divergent in
the uv if $d=4$ (i.e., they have pole terms at $d=4$). In other words, if
$d<4$, then the uv singularities of the (static and dynamic) bulk
theory are absorbed by the mass shift $\delta m^2$.

\subsubsection{Static surface renormalization functions}
\label{sec:statsurfrfs}

In order to generalize the above approach (RS$_2$) to the
semi-infinite case, we set $c_0=\infty$ and consider the analogs of
the $G$ functions (\ref{eq:Gfuncdef}) that result from these when the
boundary operators $\tilde{\bm{\Phi}}_{\mathcal{B}}$ and
$\bm{\phi}_{\mathcal{B}}$ are replaced by the normal
derivatives $\partial_n\tilde{\bm{\Phi}}$ and $\partial_n\bm{\phi}$,
respectively. We denote these functions as
\begin{eqnarray}
  \label{eq:Ginfdef}
  \lefteqn{ G_\infty^{(\tilde{N},N;\tilde{M},M)}}&&\nonumber\\
&\equiv&\biggl\langle \prod_{j=1}^{\tilde{N}}
\tilde{\Phi}^{\tilde{\alpha}_j}
\prod_{k=1}^N \phi^{\alpha_k}
\prod_{m=1}^{\tilde{M}}
\partial_n\tilde{\Phi}^{\tilde{\beta}_m}
\prod_{n=1}^M \partial_n\phi^{\beta_n}
\biggr\rangle^{\text{cum}}_{c_0=\infty}.\;\;\;\;
\end{eqnarray}
Following Refs.~\onlinecite{DS94} and \onlinecite{DS98}, we introduce
the static surface renormalization factor $Z_{1,\infty}(u)$ and the
renormalized surface operator $(\partial_n\bm{\phi})_{\text{ren}}$ via
\begin{equation}
  \label{eq:dnphiren}
  \partial_n\bm{\phi}=
  {[Z_\phi(u)\,
    Z_{1,\infty}(u)]}^{1/2}\,(\partial_n\bm{\phi})_{\text{ren}}\;. 
\end{equation}

Next we take the normal derivative on both sides of the
fluctuation-dissipation relation (\ref{eq:FDT}) with respect to
$\bm{x}'$ and set $t'=0$. The result reads
\begin{equation}
  \label{eq:normFDT}
   -\theta(t)\,{\big\langle\dot{\phi}^\alpha({\bm{x}},t)\,\partial_n{\bm{\phi}}^\beta ({\bm{x}}_{\mathcal{B}},0)\big\rangle}
=\big\langle{\phi}^\alpha({\bm{x}},t)\,
\partial_n\tilde{\Phi}^\beta({\bm{x}}_{\mathcal{B}},0)\big\rangle\,,
\end{equation}
which suggests to renormalize $\partial_n\tilde{\bm{\Phi}}$ in
complete analogy with Eq.~(\ref{eq:dnphiren}) by
\begin{equation}
  \label{eq:dnPhiren}
  \partial_n\tilde{\bm{\Phi}}=
  {[Z_\phi(u)\, Z_{1,\infty}(u)]}^{1/2}\,
  {\big(\partial_n\tilde{\bm{\Phi}}\big)}_{\text{ren}}\;.
\end{equation}
This definition ensures that the modified fluctuation-dissipation
relation (\ref{eq:normFDT}) carries over to the renormalized theory.
Moreover, it establishes consistency with the renormalization of the
corresponding static correlation function in Ref.~\onlinecite{DS98},
provided we fix $Z_{1,\infty}(u)$ as in Eqs.~(7.10a--10b) of that
reference. To this end we define the renormalized $G_\infty$ functions
via
\begin{equation}
  \label{eq:Gfuncmassren}
   G_{\infty,\text{ren}}^{(\tilde{N},N;\tilde{M},M)}=
   Z_\phi^{-\frac{\tilde{N}+N+\tilde{M}+M}{2}}\,Z_{1,\infty}^{-\frac{\tilde{M}+M}{2}}\,
   G_\infty^{(\tilde{N},N;\tilde{M},M)}\;, 
\end{equation}
if $(\tilde{N},N;\tilde{M},M)\ne(0,0;1,1)$. The excluded function
\begin{eqnarray}
  \label{eq:G0011funcmassren}
  \lefteqn{\hat{G}^{(0,0;1,1)}_{\infty,\text{ren}}(\bm{p},\omega)}&&
\nonumber\\&=&
[Z_\phi\,Z_{1,\infty}]^{-1}\,\Big[\hat{G}^{(0,0;1,1)}_\infty(\bm{p},\omega)
\mbox{}-\hat{G}^{(0,0;1,1)}_\infty(\bm{0},0)\Big]\;,\nonumber\\
\end{eqnarray}
requires an additive counterterm, which we choose such that the
normalization condition
\begin{equation}
  \label{eq:massaddct}
  \left.
   \hat{G}^{(0,0;1,1)}_{\infty,\text{ren}}(\bm{p},\omega)\right|_{p=\omega=0}=0
\end{equation}
holds. To specify $Z_{1,\infty}$, we require that
\begin{equation}
  \label{eq:Z1inffix}
  \left.\frac{\partial}{\partial p^2}\,
   \hat{G}^{(0,0;1,1)}_{\text{ren}}(\bm{p},\omega)\right|_{p=\omega=0}=
   -\frac{1}{2m}\;.
\end{equation}

Equations (\ref{eq:massaddct}) and (\ref{eq:Z1inffix}) are equivalent
to the normalization conditions (7.10a,b) of Ref.~\onlinecite{DS98};
together with Eq.~(\ref{eq:G0011funcmassren}) they imply that $G^{(0,0;1,1)}$
requires a subtraction and the renormalization factor $Z_{1,\infty}$ is
the same as in the static case.

\begin{widetext}
\subsection{Callan-Symanzik equations}\label{sec:CSE}

The Callan-Symanzik (CS) equations can now be derived in a standard
fashion. We take a derivative $\partial_{\tau_0}$ of the bare
(dimensionally regularized) $G_\infty$ functions (\ref{eq:Ginfdef}) at
fixed values of the other bare interaction constants $u_0$ and
$f_0$. Using the above reparametrizations and definitions of the beta and
exponent functions then yields
\begin{equation}
  \label{eq:CSE}
  \left[{\mathcal{D}}_m
+\frac{\tilde{N}+N+\tilde{M}+M}{2}\,\eta_\phi+\frac{\tilde{M}+M}{2}\,\eta_{1,\infty}
\right]G_{\infty,\text{ren}}^{(\tilde{N},N;\tilde{M},M)}=
R_{\infty,\text{ren}}^{(\tilde{N},N;\tilde{M},M)}
\end{equation}
with
\begin{equation}
  \label{eq:Dm}
  {\mathcal{D}}_m=m\frac{\partial}{\partial m}+\beta_u\frac{\partial}{\partial u}+\beta_f\frac{\partial}{\partial f}
-(4-\eta_\lambda)\,\lambda{\partial_\lambda}
\end{equation}
and
\begin{eqnarray}
  \label{eq:Rren}
  R_{\infty,\text{ren}}^{(\tilde{N},N;\tilde{M},M)}&\equiv&
 Z_\phi^{-(\tilde{N}+N+\tilde{M}+M)/2}Z_{1,\infty}^{-(\tilde{M}+M)/2}\,(m
 \left.\partial_m\right|_0\tau_0)\,\partial_{\tau_0}
G_\infty^{(\tilde{N},N;\tilde{M},M)}\nonumber\\&=&
(2-\eta_\phi)\,m^2\,\big[\partial_{\tau_0}
G_{\infty}^{(\tilde{N},N;\tilde{M},M)}\big]_{\text{ren}}\;.
\end{eqnarray}
Here the exponent function $\eta_{1,\infty}(u)$ is defined by setting
$\kappa=(1,\infty)$ in Eq.~(\ref{eq:etalambdamassdef}). Just like the
other static functions $\eta_u(u)$, $\eta_\phi(u)$,
$\eta_{\phi^2}(u)$, it depends only on $u$ (and $d$), but not on the
dynamic coupling constant $f$, and is precisely the same as in
Ref.~\onlinecite{DS98}. The inhomogeneities
$R_{\infty,\text{ren}}^{(\tilde{N},N;\tilde{M},M)}$ involve
renormalized functions with an insertion of $-\int d^dx\,\phi^2/2$,
given by
\begin{equation}
  \label{eq:Gparttauren}
\big[\partial_{\tau_0}
G_{\infty}^{(\tilde{N},N;\tilde{M},M)}\big]_{\text{ren}}=
Z_\phi^{-(\tilde{N}+N+\tilde{M}+M)/2}Z_{1,\infty}^{-(\tilde{M}+M)/2}
Z_{\phi^2}\partial_{\tau_0}G_{\infty}^{(\tilde{N},N;\tilde{M},M)}\;.
\end{equation}
\end{widetext}

We proceed along lines similar to those followed, for example, in
Refs.~\onlinecite{BB81} and \onlinecite{ZJ96}, in order to derive
the asymptotic scaling forms of the response and correlation
functions from the CS equations (\ref{eq:CSE}). The deviation $\delta\tau_0$
of the bare variable $\tau_0$ from its bulk critical value $\tau_{0\text{c}}$
depends on the temperature difference $\tau\equiv (T-T_{\text{c}})/
T_{\text{c}}$ according to
\begin{equation}
  \label{eq:deltatau}
  \delta\tau_0\equiv\tau_0-\tau_{0\text{c}}\sim\tau\;,
\end{equation}
which holds if $\tau$ is sufficiently small. Near criticality the mass
$m$---i.e., the inverse $\xi^{-1}$ of the (second-moment) bulk
correlation length $\xi$---behaves as
\begin{equation}
  \label{eq:corrlength}
  m\sim |\tau_0-\tau_{0\text{c}}|^\nu\quad\text{with}\quad
\nu=(2+\eta^*_{\phi^2})^{-1}\;.
\end{equation}
Furthermore, integration of Eqs.~(\ref{eq:etalambdamassdef}) and
(\ref{eq:betafmassdef}) gives the asymptotic dependence
\begin{equation}
  \label{eq:Zkappam}
  Z_\kappa\sim  m^{\eta_\kappa^*}\,,
  \quad\kappa=\phi,\,\phi^2,\,\lambda,\,(1,\infty)\,,
\end{equation}
for $(u,f)\to (u^*,f^*)$ or $m\to 0$. We insert this result for
$\kappa=\lambda$ into Eq.~(\ref{eq:lambdaren}), substitute expression
(\ref{eq:etalambdamassstar}) for $\eta^*_\lambda$, and arrive at
\begin{equation}
  \label{eq:lambdam}
  \lambda\sim m^{\mathfrak{z}}\,\lambda_0\;,
\end{equation}
where $\mathfrak{z}$ is the dynamic bulk exponent
(\ref{eq:dynz}). With the aid of these results, it is straightforward
to deduce the scaling forms
\begin{eqnarray}
  \label{eq:scfGs}
  \lefteqn{G_\infty^{(\tilde{N},N;\tilde{M},M)}(\{\bm{x}\},t;\tau_0,u_0,f_0,\lambda_0)}&&\nonumber\\
&\approx&
  m^{(\tilde{N}+N)\beta+(\tilde{M}+M)\beta_1^{\text{ord}}}\,
\Xi_\infty^{(\tilde{N},N;\tilde{M},M)}(\{m\bm{x}\},\lambda_0\,tm^{\mathfrak{z}})\;.
\nonumber\\
\end{eqnarray}
Here $\beta=(\nu/2)(d-2+\eta)$ is a standard bulk critical index,
while $\beta_1^{\text{ord}}$ is its surface counterpart for the
\emph{ordinary} transition. (For recent estimates of the numerical
value of the latter at $d=3$, see Ref.~\onlinecite{DS98}, its
references, and Ref.~\onlinecite{MKslab}.) The set $\{\bm{x}\}$
comprises all position coordinates on which the respective function
depends. The case $(\tilde{N},N;\tilde{M},M)=(0,0;1,1)$ is special in
that the term $\delta(t{-}\tilde{t})\,
\delta({\bm{x}}_\|{-}\tilde{\bm{x}}_\|)\,
\hat{G}^{(0,0;1,1)}({\bm{p}{=}\bm{0},\omega{=}0})$, which results from
the subtraction in Eq.~(\ref{eq:G0011funcmassren}), should be
subtracted on the left-hand side of Eq.~(\ref{eq:scfGs}). We have
suppressed this term, because we consider $G^{(0,0;1,1)}$ here not as
a distribution, but as a function for $t-\tilde{t}>0$.

Let us choose $(\tilde{N},N;\tilde{M},M)=(0,2;0,0)$ in
Eq.~(\ref{eq:scfGs}) and consider the case of the spin-spin cumulant
(\ref{eq:Sijtt}). If we set $\lambda_0=1$ for convenience, we obtain the
scaling form given by Eq.~(\ref{eq:Czzrt}) in the limit $m\to 0$.

The scaling form of the surface structure function (\ref{eq:scf}) at the
ordinary transition can be derived from the expansion of $G^{(0,0;0,2)}$
in powers of $c_0^{-1}$ (see Eq.~(\ref{eq:scfGs})). Alternatively, we
can combine the CS equations (\ref{eq:CSE}) with Eq.~(\ref{eq:scfGs})
and the BOE (\ref{eq:boephi}) (applied to the renormalized theory) to
conclude that the coefficient function $D(z,t)$ asymptotically
satisfies the CS equation
\begin{equation}
  \label{eq:RGC}
 {\left[
   {\mathcal{D}}_m-\frac{\eta_{1,\infty}}{2}\,\right]}\,D(z,t) = 0\;.
\end{equation}
In the limit $m\to 0$, Eq.~(\ref{eq:RGC}) yields a leading short-distance
singularity of the form
\begin{equation}
  \label{eq:Dz}
  D(z,t)\mathop{\approx}\limits_{z\to 0} D_0\,z^{1+\eta_{1,\infty}^*/2}\;,
\end{equation}
where the exponent $\eta_{1,\infty}^*$ can be expressed in terms of $\eta$
and the surface correlation index
$\eta^{\text{ord}}_\|=2+\eta+\eta_{1,\infty}^*$. It follows that the
scaling function $\Upsilon$ in Eq.~(\ref{eq:Czzrt}) must behave as
\begin{equation}
  \label{eq:BOEC}
  \Upsilon\left({\mathsf{z}},{\mathsf{z}}';{\mathsf{t}}\right)
\mathop{\approx}\limits_{{\mathsf{z}},{\mathsf{z}}'\to 0}
\left({{\mathsf{z}}{\mathsf{z}}'}
\right)^{(\eta_{\|}^{\text{ord}}-\eta)/2}
\,
{\Upsilon_0}({\mathsf{t}})\;.
\end{equation}
This in turn implies that the Fourier transformed surface structure
function $\hat{C}_{11}(\bm{p},\omega)$ at $T_{\text{c}}$ can be written as
\begin{equation}\label{eq:C11}
{\hat{C}_{11}}(\boldsymbol{p},\omega) \approx
p^{\eta_{\|}^{\text{ord}}-1-{\mathfrak{z}}}\, {\sigma}{\left(\omega
p^{-{\mathfrak{z}}}\right)}\;.
\end{equation}
The limit $\bm{p}\to\bm{0}$ exists. By consistency, we must therefore have
\begin{equation}
\label{eq:C0}
{\hat{C}_{11}}(\boldsymbol{0},\omega) = \text{const}\;
\omega^{-{\left({\mathfrak{z}}+1-\eta_{\|}^{\text{ord}}\right)}/{\mathfrak{z}}}\;.
\end{equation}

In the next section, we will check the predictions (\ref{eq:C11})
and (\ref{eq:C0}) by means of accurate Monte-Carlo spin dynamics
simulations.
 
\section{Monte-Carlo spin dynamics simulation}
\label{sec:MC}

The Monte-Carlo spin dynamics simulation works as follows: A
Monte-Carlo simulation of the lattice model with the Hamiltonian 
(\ref{eq:Hamil}) yields a spin configuration that is used as
initial condition for the integration of the equations of motion
(\ref{eq:eqmot}). When the integration is completed, the time
evolution of the spin configuration is analyzed for position and time
displaced correlations. The correlation functions are then stored in
arrays, and a new initial condition is generated by the Monte-Carlo
simulation.  Typically, this is repeated 700 to 1000 times. The
correlation function $C^{\alpha \beta}{\left(\boldsymbol{r}; z, z';
|t {-} t'| \right)}$ [cf.\ Eq.~(\ref{eq:Sijtt})] is finally obtained by
averaging over the individual measurements.

The Monte-Carlo algorithm is chosen as a hybrid scheme consisting
of Metropolis sweeps, Wolff single cluster updates,\cite{Wol89} and
overrelaxation sweeps.\cite{CFL93} Typically, one hybrid Monte-Carlo
step involves 10 individual steps, each of which can be one of the
updates listed above. Both the Metropolis and the Wolff algorithm work
in the standard way, where the reduced coordination number of the
lattice at the surfaces and the modified surface coupling $J_1$ must
be taken into account.  The acceptance probability $p$ of a proposed
spin flip in the Metropolis algorithm is defined by $p(\Delta E) =
1/[\exp(\Delta E / k_B T) + 1]$, where $k_B$ is Boltzmann's constant
and $\Delta E$ is the change in configurational energy of the proposed
move.

The overrelaxation part of the algorithm performs a microcanonical
update of the configuration by sequentially rotating each spin in the
lattice such that its contribution to the energy of the whole
configuration remains constant. The implementation of this update
scheme is straightforward. To see this, note that the energy of a spin
with respect to its neighborhood has the functional form of a scalar
product according to Eq.~(\ref{eq:Hamil}). Therefore, a spin can be
rotated about the direction of the local field generated by its
neighbors without changing the local energy. The angle of rotation can
be chosen randomly for each spin. However, in order to have
minimal autocorrelation times, a reflection---i.e., a
rotation of a spin by 180 degrees---turns out to be the most
efficient overrelaxation move. In one overrelaxation sweep this update
is applied in sequence to every spin of the lattice.

Typically, a hybrid Monte-Carlo step consists of two sweeps of the
whole lattice via the Metropolis algorithm (M), four sweeps of the
whole lattice by means of the overrelaxation algorithm (O) described
above, and four single cluster updates according to the Wolff
algorithm (C). The individual updates are mixed automatically in the
program so as to generate the update sequence M\ O\ C\ O\ C\ M\ O\ C\ 
O\ C\ . As our random number generator, we have utilized the
shift-register generator R1279 defined by the recursion relation $X_n
= X_{n-p} \oplus X_{n-q}$ for $(p,q) = (1279,1063)$. Generators like
this one are among the fastest available. However, they are known to
cause systematic errors in combination with the Wolff
algorithm.\cite{cluerr} For lags $(p,q)$ as large as the ones used
here, these errors are much smaller than typical statistical errors.
The hybrid nature of the algorithm reduces them further.\cite{AFMDPL}

Using this Monte-Carlo scheme, we have investigated lattice sizes $L$
between $L = 12$ and $L = 72$. The integrated correlation time of the
hybrid algorithm is determined by the autocorrelation function of the
energy or, equivalently, by the autocorrelation function of the
modulus of the magnetization. Either quantity is $O(3)$ symmetric, and
for sufficiently long times, the decay of the corresponding
autocorrelation functions is governed by the same autocorrelation
time. This time scale characterizes the slowest mode of the Wolff
algorithm, so it also determines the correlation time of our
hybrid Monte-Carlo algorithm.  Note that the autocorrelation time of
the Metropolis algorithm is determined by the decay of the
autocorrelation function of the {\em order parameter}, which decays
particularly slowly near the critical point (critical slowing down). For
the hybrid scheme described above, the autocorrelation time does not
exceed 10 hybrid Monte-Carlo steps for the largest lattice size at $T
= T_{\text{c}}$.  In order not to obtain too strongly correlated
initial conditions for the equation of motion, an initial condition is
generated every tenth hybrid Monte-Carlo step.

The integration procedure for the equations of motion is completely
separated from the Monte-Carlo part of the program. The second-order
sublattice decomposition integrator described in Ref.~\onlinecite{SD}
is used here. Long-time stability is provided by the exact conservation of
energy [see Eq.~(\ref{eq:eqmot})] and spin normalization. The magnetization
is only conserved within the accuracy of the discretization, i.e., to
second order in the time step. Typical time steps $\delta t$ used here
range from $\delta t = 0.04/J$ to $\delta t = 0.08/J$, depending on the
size of the system. For the largest system $(L = 72)$, the total integration
time is $\tau_I = 8192/J$; in this case $\delta t = 0.08/J$
was used. Note that the decomposition integrator $I(\delta t)$ has the
exact time inversion property $I(-\delta t) = I^{-1}(\delta t)$. This
guarantees that the time evolution of discretization errors, such as
those affecting the conservation of the magnetization, does not contain
systematic drifts.\cite{SD}

If the algorithm is implemented on a single processor machine, the
major part of the CPU time is consumed by the integration of the
equations of motion. This fraction increases with system size because
the CPU time needed for the integration scales as $\tau_I L^3$,
whereas that of a hybrid Monte-Carlo step scales as $L^3$. If Wolff
updates are used exclusively, the average scaling is reduced to
$L^{2-\eta}$.\cite{Wol89} For the purposes of the present
investigation the integration time $\tau_I$ has to be chosen such that
the slowest spin wave or spin diffusion modes in the system can be
identified.  At $T = T_{\text{c}}$ this means that $\tau_I \sim
L^{\mathfrak{z}}$ [see Eq.~(\ref{eq:dynz})]. Below $T_{\text{c}}$, one
must have $\tau_I \sim L^2$ for an isotropic ferromagnet in order to
resolve the slowest spin-wave mode. Above $T_{\text{c}}$, the dynamics
is dominated by spin diffusion, which also requires $\tau_I \sim L^2$
for the resolution of the slowest modes.  It is therefore very
desirable to distribute the integration task of the simulation over
several processors on a parallel machine.

A simple and very efficient implementation on a parallel machine with
at least four processors can be constructed according to the following
master-slave idea: The master process runs the Monte-Carlo part of the
simulation to provide initial conditions on demand, and the slave
processes integrate the equations of motion for different initial
conditions in parallel and analyze the correlations. The amount of
communication among the processes is determined by the transfer of
initial configurations from the master to the slaves at the beginning
of the simulation, and by the transfer of the correlation data from
the slaves to the master for the final evaluation and output. If $N$
processors in parallel are used in this way, the speedup is very close
to the theoretical limit $N-1$ for sufficiently large integration
times $\tau_I$. We have implemented the master-slave version of the
spin dynamics algorithm on the Alpha Linux cluster ALiCE at the {\em
  Institut f\"ur angewandte Informatik} at the Bergische
Universit{\"a}t Wuppertal, using up to 32 processors in parallel for
the largest systems. Communication between the processors is
facilitated by the MPI (message passing interface) communication
library.

A well-known major problem one is faced with in any computer
simulation study of critical behavior is how to extract the asymptotic
critical behavior from the data. This is particularly challenging in
our case since we have to cope with two additional complications:
\emph{surface} critical behavior and \emph{dynamics}. In the
asymptotic critical regime the value of the ratio $r_1\equiv J_1/J$
does not matter if $d=3$ because surfaces of three-dimensional
isotropic Heisenberg ferromagnets are always disordered in the absence
of external fields. Thus such systems always belong to the ordinary
surface universality class. However, to what extent the asymptotic
scaling can actually be observed in a computer simulation on finite
systems is a completely different issue. 

The experience made in a previous study of the static case by one of
us\cite{MKslab} suggests that it should be possible to avoid extended
crossover regimes by a careful choice of the ratio $r_1$. In order to
find out which value of $r_1$ is optimal in the sense of giving the
largest asymptotic regime, we proceed as in Ref.~\onlinecite{MKslab}:
We consider the magnetization profile in thermal equilibrium,
determine $r_1$ in such a way that a discrete version of the Dirichlet
boundary condition holds, and then verify that this choice suppresses
corrections to the asymptotic behavior, making the asymptotic regime
larger than for alternative values of $r_1$. Let us explain this in
detail. The equilibrium profile is
\begin{equation}
\label{eq:mofz}
m(z) \equiv \left\langle \frac{\bm{M}_{\text{tot}}}
{|\bm{M}_{\text{tot}}|} \cdot \sum_{i_1,i_2=0}^{L-1}
\bm{S}_{(i_1,i_2,i_3)} \right\rangle\;,
\end{equation}
where $\bm{M}_{\text{tot}} \equiv \sum_{\bm{i}} \bm{S}_{\bm{i}}$ is the
total magnetization, while $z \equiv i_3 + 1/2$ with $i_3 = 0, \dots,
L-1$ indicates a lattice plane parallel to the surfaces [cf.\ 
Eq.~(\ref{eq:Hamil})]. Note that according to this definition of $z$,
the `boundary planes' $z=1/2$ and $z=L-1/2$ of the system are located
half a lattice constant away from the first and last lattice layers,
respectively. With this convention, the lattice model defined by
Eq.~(\ref{eq:Hamil}) may be viewed as a {\em discrete version} of the
Ginzburg-Landau Hamiltonian (\ref{eq:HGL}), where the order parameter of
the numerical cell $\bm{i}$ is represented by the spin
$\bm{S}_{\bm{i}}$ at its \emph{center}.

The bulk magnetization $m_{\text{b}}$ can be approximated by the value
of the magnetization in the center layer of the system, i.e.,
$m_{\text{b}} \equiv m(z_{\text{mid}})$ with $z_{\text{mid}} = L/2$
when the number of layers, $L$, is odd. For even $L$,
$z_{\text{mid,1}}=(L-1)/2$ and $z_{\text{mid,2}}=(L+1)/2$ are
equivalent choices for $z_{\text{mid}}$; in this case, $m_{\text{b}}
\equiv (m(z_{\text{mid,1}}) + m(z_{\text{mid,2}}))/2$ is interpreted
as the bulk magnetization. The ratio $m(z)/m_{\text{b}}$ then depends
only on $z/L$, which motivates us to define the scaling function
\begin{equation}
\label{mscal}
M(z/L) \equiv m(z) / m_{\text{b}}\;,
\end{equation}
where $T = T_{\text{c}}$ is assumed. The analysis of the data reveals that
the scaling function $M(\zeta \equiv z/L)$ can be represented by the
simple fit formula
\begin{equation}
\label{eq:mzfit}
M(\zeta) = B_M \left[ (\zeta + \zeta_0)
(1 - \zeta + \zeta_0) \right]^{(\beta_1 - \beta)/\nu}
\end{equation}
to a remarkable accuracy. Here $\zeta_0 = z_0/L$ is the scaled
extrapolation parameter.  In analyzing the data we have accepted the
estimates $\beta =0.3662\pm 0.0025$ and $\nu =0.7073\pm 0.0035$ of
Ref.~\onlinecite{GZJ98}, and utilized the value $\beta_1 = 0.834(6)$
of Ref.~\onlinecite{MKslab}.

From a least square fit of Eq.~(\ref{eq:mzfit}) to the data for
various system sizes we obtain the extrapolation parameter $z_0$ in
units of the lattice spacing. For the choices $J_1/J = 0.3$ and 1.0,
we find $z_0 \simeq -0.34$ and $z_0 \simeq 0.46$, respectively. On the
other hand, $z_0$ \emph{vanishes} within the statistical errors if
$J_1/J = 0.73$.\cite{MKslab} To put these findings in perspective,
some explanations are necessary. Owing to our definition of $z$ [given
below Eq.~(\ref{eq:mofz})], a fit curve (\ref{eq:mzfit}) with scaled
extrapolation parameter $\zeta_0=0$ means that the measured
magnetization profile extrapolates to zero half a lattice constant
away from the outermost layers $i_3 = 0$ and $i_3 = L-1$ of our
lattice model. In this sense, the profile satisfies a Dirichlet
boundary condition \emph{on the scale of the lattice constant} in this
special case.

Let us emphasize that such a boundary condition on a microscopic scale
\emph{must not be confused} with the Dirichlet boundary condition
which the order parameter satisfies at the ordinary transition
\emph{on long scales}, irrespective of the precise value of the ratio
$r_1\equiv J_1/J$. The latter is an \emph{asymptotic long-scale
  property}, associated with the corresponding ordinary fixed point of
the RG, and hence \emph{universal}. By contrast, the boundary
condition that the order-parameter profile of a given microscopic
model is found to satisfy on microscopic scales generally
\emph{depends on microscopic details}, and is therefore a
\emph{nonuniversal property} (cf.\ the discussion in Sec.~III.C.9 of
Ref.~\onlinecite{Diehl}).

On the level of a mesoscopic description via our continuum model, a
Dirichlet boundary condition can be enforced on the mesoscopic length
scales on which such a description makes sense (several lattice
constants) by setting the enhancement variable $c_0$ to the
fixed-point value $+\infty$. For values $c_0<\infty$, the Dirichlet
boundary condition does \emph{not} hold on mesoscopic scales, neither
for the regularized nor for the renormalized theory. In other words, a
$c_0$-dependent extrapolation parameter $z_0\ne 0$ occurs. This
deviation from the Dirichlet boundary condition corresponds to a
correction to scaling: It is irrelevant inasmuch as it vanishes in the
limit $z_0/z\to 0$.  Choosing a particular value
$r_1=r_1^{(\text{D})}$ for the ratio of interaction constants such
that $z_0\simeq 0$ is an appealing way to mimic the Dirichlet boundary
condition of the $c_0=\infty$ continuum theory on a lattice. As we
know already for the static case from Ref.~\onlinecite{MKslab}, and
will verify for the dynamic theory below, this choice of $r_1$
suppresses corrections to scaling and hence enlarges the regime in
which the asymptotic scaling behavior is observed.

The optimal value which yields a vanishing extrapolation parameter
$z_0$ for temperatures sufficiently close to $T_{\text{c}}$ and
moderately large lattice dimensions $L$ is $r_1^{(D)}=0.73$. For
values close to this optimal one, we have
\begin{equation}
  \label{eq:z0}
  z_0 = a\ (r_1 - r_1^{(D)}) + {\mathcal{O}}[(r_1 - r_1^{(D)})^2]\;,
\end{equation}
where $a$ is a factor of order unity. This behavior was already
obtained in Ref.~\onlinecite{MKslab}, where crossover effects and the
behavior of the order parameter profile as a function of $r_1$ were
investigated in more detail for the static case. The present work
confirms these findings: Our results for the dynamic surface structure
function presented in the next section are fully consistent with them.

\section{The dynamic surface structure function}\label{sec:dynssf}

\begin{figure}[htbp]
\includegraphics[height=\columnwidth,angle=-90]{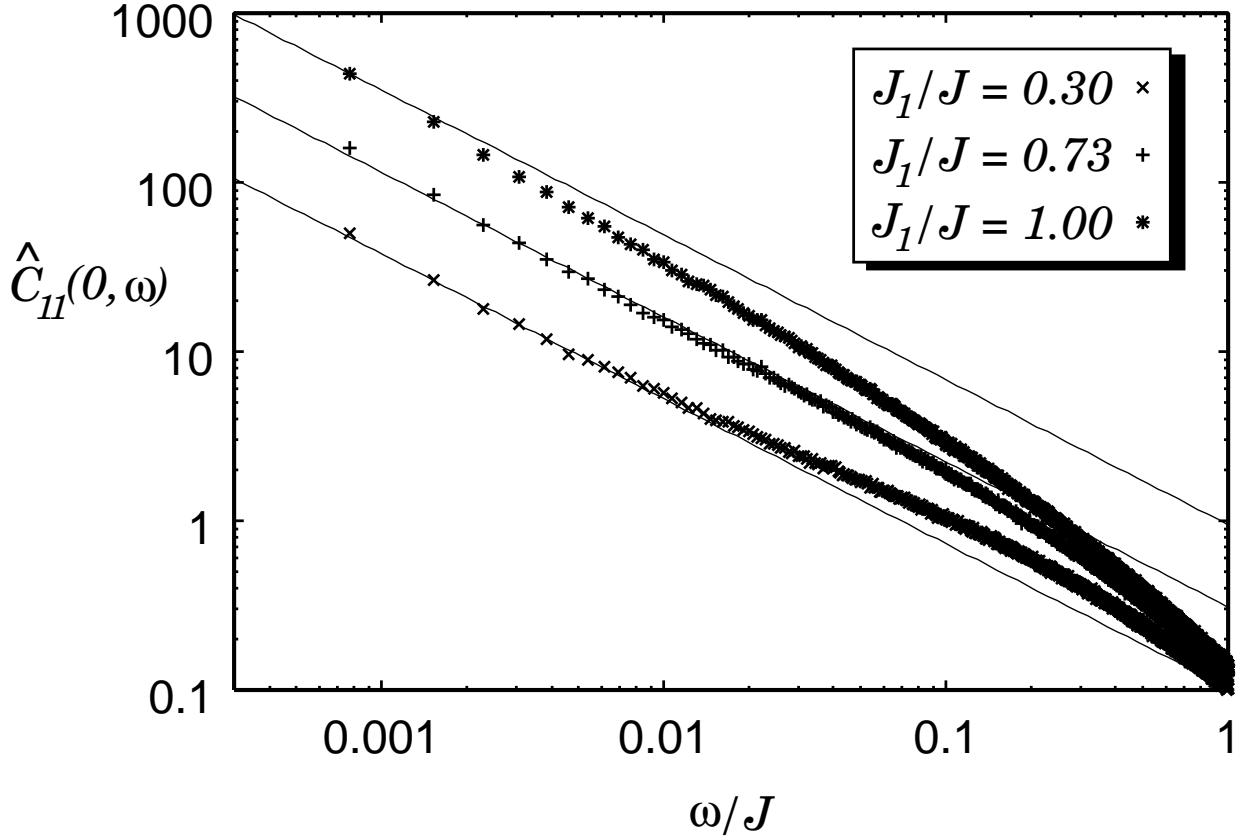}
\caption{Structure function
  $\hat{C}_{11}(\bm{0},\omega)$ for $r_1\equiv J_1/J$ = 0.3
  $(\times)$, 0.73 $(+)$, and 1.0 $(*)$.  Error bars (one standard
  deviation) are smaller than the symbol sizes.  The solid lines
  indicate the theoretically expected power law (\ref{eq:C0}) for
  $\omega \to 0$.
\label{S10wfig}}
\end{figure}

\begin{figure}[htbp]
\includegraphics[height=\columnwidth,angle=-90]{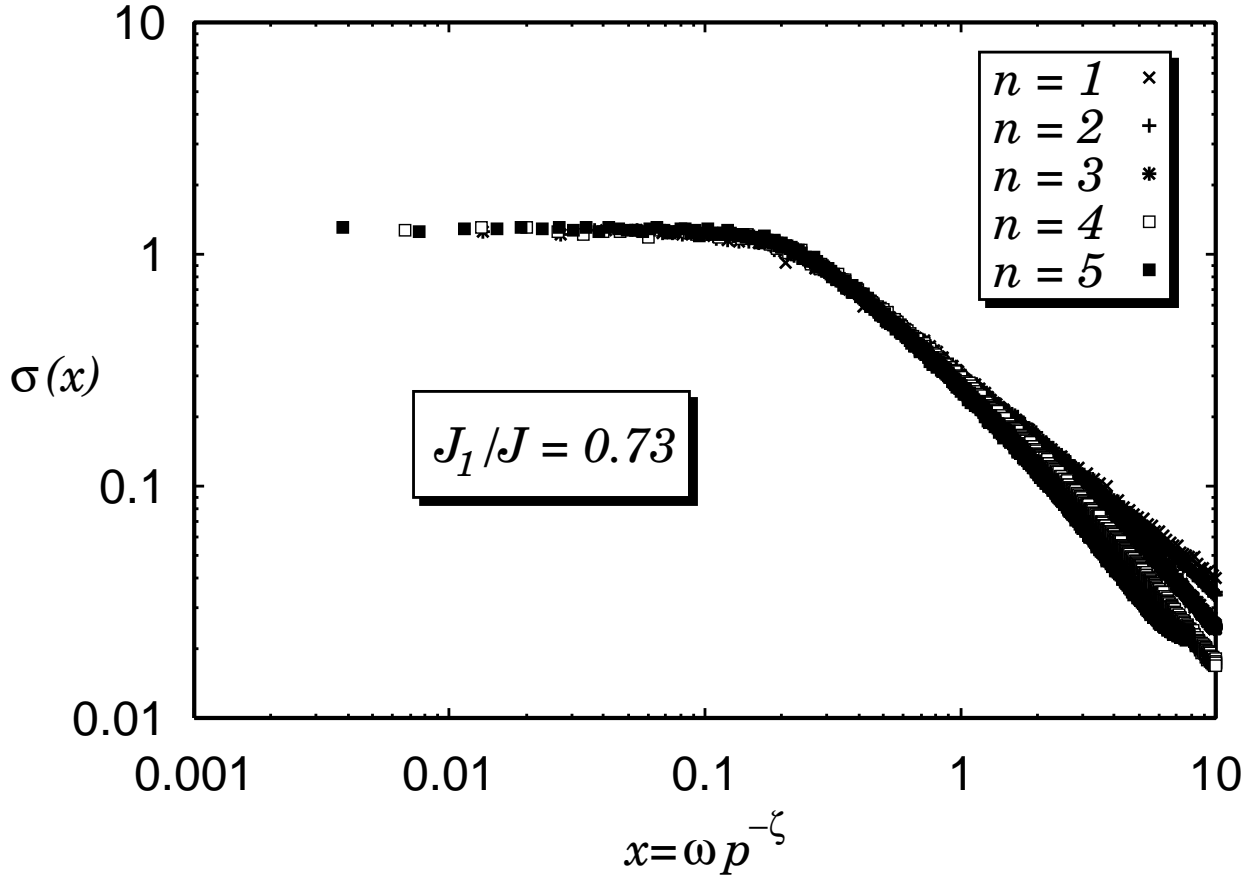}
\caption{Scaling function $\sigma(x)$ according to
  (\ref{eq:C11}). Data obtained for $r_1\equiv J_1/J = 0.73$ and
  $\bm{p}=(\frac{n\pi}{36},0,0)$, with $n= 1,\ldots,5$, are shown.
  Error bars (one standard deviation) are smaller than the symbol
  sizes. The data follow Eq.~(\ref{eq:sigma}) up to $x \simeq 1$. The
  data for $x \geq 1$ are outside the scaling regime.
\label{S1pw}}
\end{figure}

\begin{figure}[htbp]
\includegraphics[height=\columnwidth,angle=-90]{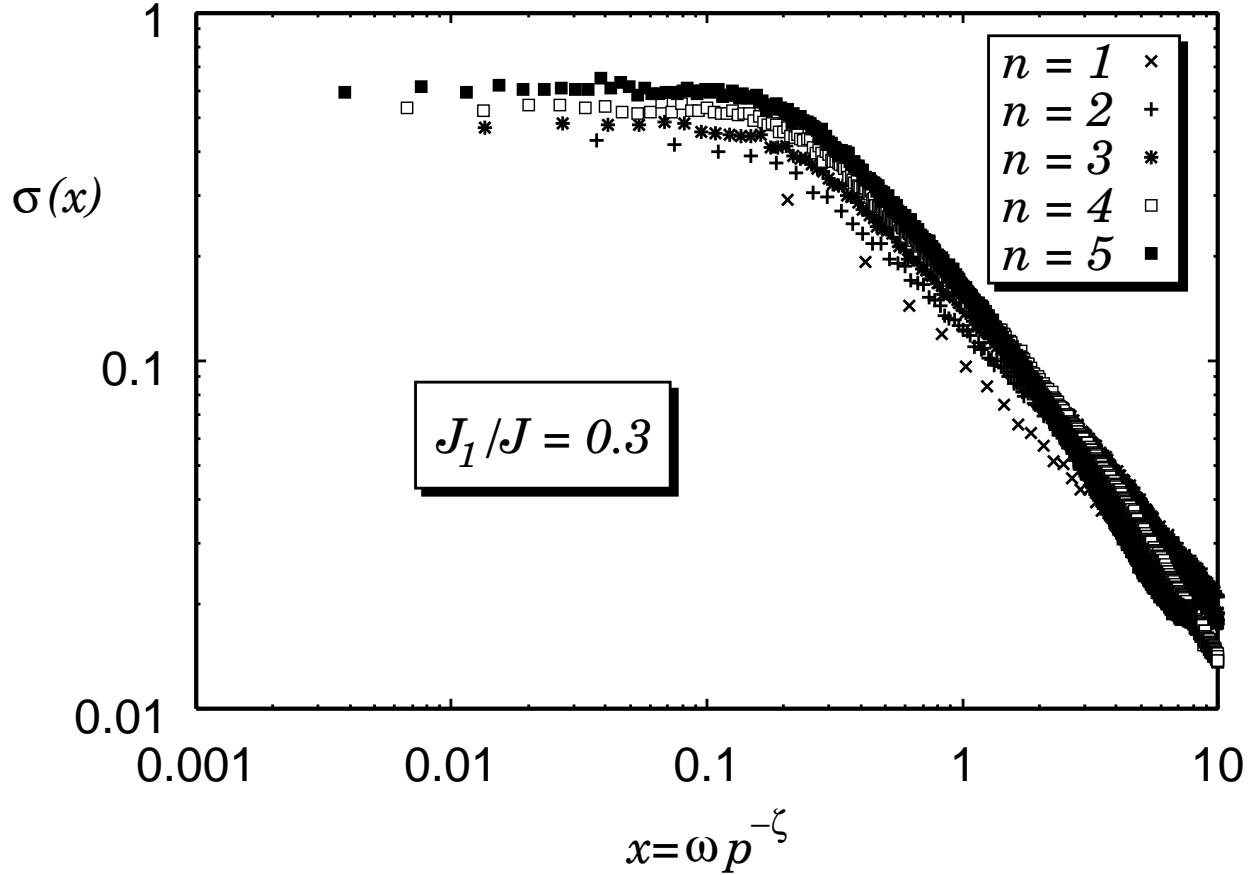}
\caption{Scaling plot of the surface structure function for $r_1\equiv
  J_1/J = 0.3$ and $\bm{p}=(\frac{n\pi}{36},0,0)$, with $n=
  1,\ldots,5$.  Error bars (one standard deviation) are smaller than
  the symbol sizes. The data do not follow Eq.~(\ref{eq:sigma}).
\label{S1pw3}}
\end{figure}

\begin{figure}[htbp]
\includegraphics[height=\columnwidth,angle=-90]{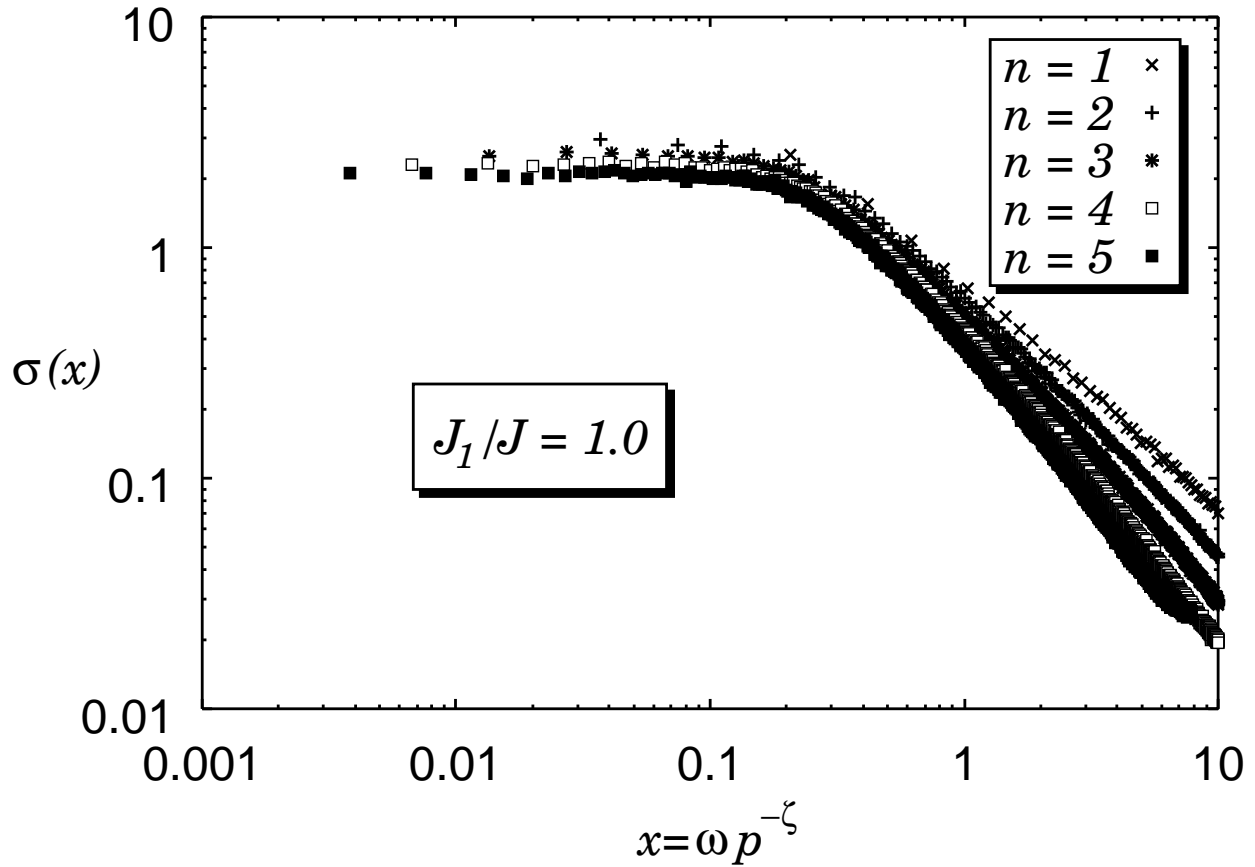}
\caption{Scaling plot of the surface structure function for $r_1 = 1.0$
  and $\bm{p}=(\frac{n\pi}{36},0,0)$, with $n= 1,\ldots,5$.  Error bars
  (one standard deviation) are smaller than the symbol sizes. The data
  do not follow Eq.~(\ref{eq:sigma}).
\label{S1pw10}}
\end{figure}

Our simulation results are displayed in
Figs.~\ref{S10wfig}--\ref{S1pw10}. They were obtained for $T =
T_{\text{c}}$, $L = 72$, and the  total integration time $\tau_I =
8192/J$.  For the smallest accessible frequency $\omega_{\text{min}} =
2\pi / \tau_I$, finite-size effects turned out to be negligible.

In Fig.~\ref{S10wfig} the structure function
$\hat{C}_{11}(\bm{0},\omega)$ is shown for different values of
$r_1$ in comparison with Eq.~(\ref{eq:C0}). The exponent
$({\mathfrak{z}} + 1 - \eta_{\|}^{\text{ord}})/{\mathfrak{z}}$
has the value $0.856 \pm 0.005$ that follows from the estimate
${\mathfrak{z}}(d{=}3)=2.482\,{\pm}\,0.002$ obtained by the substitution
$\eta(d{=}3)=0.036\pm 0.004$ \cite{GZJ98} in Eq.~(\ref{eq:dynz})
and the current estimate of the surface correlation exponent
$\eta^{\text{ord}}_{\|}(d{=}3)=1.358 \pm 0.012$ \cite{MKslab}
of the ordinary transition.

The dependence of $\hat{C}_{11}(\bm{0},\omega)$ on $r_1$ is
particularly interesting. If $r_1$ is small $(r_1=0.3, \times)$, our
simulation data approach the asymptotic power law (\ref{eq:C0}) from
\emph{above}, whereas for larger values of $r_1$ $(r_1=1, *)$, the
asymptotic power law is approached from below. In the latter case, the
data even remain \emph{outside} the asymptotic regime for the
frequency range shown in Fig.~\ref{S10wfig}. The best agreement with
Eq.~(\ref{eq:C0}) over the largest frequency range is obtained for the
choice $r_1 = 0.73 (+)$, which has already been identified as optimal
in the sense that the extrapolation parameter $z_0$ for the
magnetization profile vanishes (see Eq.~(\ref{eq:z0})). The deviations
from the power law (\ref{eq:C0}) for $r_1 \neq 0.73$ can apparently be
attributed to \emph{dynamic} surface-induced corrections to the
asymptotic behavior that originate from the nonzero, $r_1$-dependent
value of the extrapolation parameter $z_0$.

Fig.~\ref{S1pw} shows a scaling plot of $\hat{C}_{11}({\bm{p}},\omega)$,
where $\bm{p} = (\frac{n\pi}{36},0,0)$ is oriented along the surface.
As one sees, the scaling regime in $x$ shrinks as the mode index $n$
is increased from $1 (\times)$ to $5 (\rule{2mm}{2mm})$.  For $x < 1$,
the shape of the scaling function in Eq.~(\ref{eq:C11}) is described
surprisingly well by the fit function\cite{KKD01}
\begin{equation}
\label{eq:sigma}
\sigma(x) = \sigma_0 \left[ 1 + (x/x_0)^4
\right]^{(\eta^{\text{ord}}_{\|}-{\mathfrak{z}}-1)/4{ \mathfrak{z}}}\;,
\end{equation}
which is inspired by the known zero-loop result.\cite{DJ92,Die94b,WD95}
The exponent at  the square bracket is chosen so as to reproduce
Eq.~(\ref{eq:C0}) in the limit $x \to \infty$ ($p \to 0$ at fixed
$\omega\ne 0$). The amplitude $\sigma_0$ and the crossover parameter
$x_0$ are used as fit parameters.

The agreement between the data displayed in Figs.~\ref{S10wfig} and
\ref{S1pw} and the scaling forms (\ref{eq:C11}) and (\ref{eq:C0}) is
quite satisfactory. However, on closer inspection small deviations are
found to remain. Note that as pointed out at the end of the previous
section and in analogy with the results of Ref.~\onlinecite{MKslab}
for the equilibrium case, the choice $r_1 = 0.73$ yields indeed the
largest regime in which asymptotic scaling holds (see
Fig.~\ref{S10wfig}).

Hence we expect the choice $r_1 = 0.73$ to be optimal also for the
surface structure factor at \emph{finite} momentum transfer $p$. Our
results for $r_1 = 0.3$ and $r_1 = 1.0$ depicted in Figs.~\ref{S1pw3}
and \ref{S1pw10} confirm this expectation. Figure \ref{S1pw3} shows
that the data for $r_1 = 0.3$ approach the scaling function
$\sigma(x)$ of Fig.~\ref{S1pw} monotonically from \emph{below} as the
mode number $n$ is increased from 1 to 5. The nonasymptotic
surface-induced corrections are so large that the data for different
$n$ (i.e., momentum transfers ${\bf p}$) are well separated even on a
logarithmic scale. In other words, no data collapse nearly as nice as
in Fig.~\ref{S1pw} occurs, although the crossover parameter $x_0$
appears to be consistent with the results displayed there.

Our results for $r_1 =1.0$ (see Fig.~\ref{S1pw10}) show a similar
behavior, except that the scaling function $\sigma(x)$ now is
monotonically approached from \emph{above} as the mode number $n$
increases. The crossover parameter $x_0$ is again consistent with our
findings in Figs.~\ref{S1pw} and \ref{S1pw3}. The nonasymptotic
surface-induced corrections are as large as in Fig.~\ref{S1pw3} but
have opposite sign.

From these results a consistent numerical picture of dynamic surface
scaling emerges. For the optimal choice $r_1 = 0.73$, our simulation
data for the dynamic structure function bear out quite convincingly
the asymptotic behavior we predicted on the basis of our RG work. For
values of $r_1$ deviating significantly from $0.73$, the data exhibit
pronounced corrections to scaling. These entail that our data for
values like $r_1=1$ and $0.3$ do not exhibit \emph{directly} the
asymptotic scaling form and power law of the surface structure
function $\hat{C}_{11}(\bm{p},\omega)$ for nonzero and zero parallel
momentum $\bm{p}$, respectively. Yet they seem to be consistent both
with the theoretically predicted asymptotic behavior as well as with
the one extracted from our simulation data for $r_1=0.73$ because the
observed deviations appear to be attributable to corrections to
scaling.

In order to demonstrate universality, it would clearly be very
desirable to verify the approach to the asymptotic critical behavior
also for unfavorable values of $r_1$ like $1.0$ and $0.3$. One
conceivable way of trying to reach this goal is by means of
brute-force numerical means. However, in view of the enormous
numerical effort that was necessary to produce the above simulation
results, we do not consider this to be promising strategy at present.

We believe that a better strategy would be the incorporation of
nonasymptotic correction terms in the analysis of the simulation data.
Unfortunately, there are various sources of such corrections, and
detailed knowledge about their form and strength is either scarce or
not available. A systematic investigation of the various kinds of
nonasymptotic corrections of static and dynamic origin that might play
a role in the analysis of the surface critical behavior we are
concerned with here evidently requires more numerical and analytical
work, and is beyond the scope of this paper. Let us therefore simply
discuss some possible sources of deviations from scaling, beginning
with the ones that do not correspond to corrections to scaling.

Two obvious sources of this latter kind are insufficient momentum and
frequency resolution. By virtue of the relation $\delta p = 2\pi/L$
the momentum resolution $\delta p$ is intimately linked to the system
size $L$, which despite formidable progress in simulation techniques
and computational resources still is a serious limiting factor. The
frequency resolution $\delta \omega = 2\pi/\tau_I$ is limited by the
total integration time $\tau_I$.  From our data for $C_{11}(\bm{p},t)$
(not shown) we conclude that $\tau_I$ is sufficiently long. The
frequency resolution $\delta \omega / J \simeq 7.7 \times 10^{-4}$
that is available here rivals that of neutron scattering
experiments.\cite{DPLMK} The momentum resolution is given by $\delta p
\simeq 0.087$ in units of the inverse lattice constant for our largest
systems, and is therefore much more restrictive.

One familiar type of corrections to scaling are those induced by
deviations of the coupling constant $u$ from its fixed point value
$u^*$. Just as in the static case, they are governed by the Wegner
exponent $\omega_u\equiv \beta_u'(u^*)$ whose value is $\simeq 0.8$ in
$d=3$ dimensions.\cite{GZJ98} Analogous corrections to scaling result
from the RG flow of the mode-coupling interaction constant $f$. Upon
linearizing the flow about the infrared-stable fixed point
$(u^*,f^*)$, one obtains in addition to $\omega_u$ a second
correction-to-scaling exponent, $\omega_f\equiv \beta_f'(u^*,f^*)$,
which in contrast to the former is of purely dynamic origin. We are
not aware of any reliable estimates of $\omega_f$ for
$d=3$.\cite{rem:omegaf} 

Other potentially dangerous corrections might be caused by a
previously mentioned important difference of the dynamics of the
simulated lattice model and the semi-infinite continuum model J: that
the former conserves the energy while the latter does not. In Appendix
\ref{sec:ec} we generalize model J for the bulk case by incorporating
energy conservation. The resulting model is analogous to model D (with
$n=3$ components), and reduces to this when the mode-coupling
interaction constant $f$ is set to zero. We show that the ratio
$\lambda/\lambda^{({\mathcal E})}$ of transport coefficients (where
$\lambda^{({\mathcal E})}$ denotes the analog of $\lambda$ for the
energy density ${\mathcal E}$) transforms under the RG as
$\ell^{\mathfrak{z}-2}$ in the long length-scale limit $\ell\to 0$.
Since $\mathfrak{z}-2=(d-2+\eta)/2$, which is $\simeq 0.5$ in three
dimensions, the ratio approaches indeed zero, albeit with a
considerably smaller power than in the case of model D (where the
value of this exponent is $2-\eta\simeq 2$).

Hence two conclusions may be drawn: First, in order to obtain the
asymptotic critical behavior of order parameter cumulants we can take
the limit $\lambda/\lambda^{({\mathcal E})}\to 0$. If one sets
$1/\lambda^{({\mathcal E})} =0$ from the outset, the energy density
relaxes instantaneously, This means that the effects produced by the
coupling to the energy density correspond to a change of the
parameters of the original model J, up to corrections due to
irrelevant operators. In other words, the energy conservation should
not affect the asymptotic critical behavior, so that our lattice model
should belong to the universality class of our semi-infinite model J.
Second, we cannot rule out that the corrections to the asymptotic
behavior induced by the conservation of the energy are less important
for an improved analysis of the numerical data presented above than
the previously mentioned corrections to scaling. To assess the
relative importance of the various types of corrections to scaling
seems difficult without reliable additional information based on
detailed calculations.

The corrections to scaling we have just considered are associated with
irrelevant \emph{bulk} variables and hence are not specific to systems
with surfaces. Analogous ones are induced by irrelevant \emph{surface}
variables. A well-known example are the corrections $\sim \ell$
resulting from deviations of $1/c_0$ (the reciprocal surface
enhancement variable) from the fixed-point value
$1/c^*_{\text{ord}}=0$. One evident consequence of such corrections
(which is, however, not the only one when Landau theory is not valid)
is that the Dirichlet boundary condition the order-parameter density
satisfies at the ordinary fixed point ceases to hold. In view of the
two observations made above---namely, (i) that the choice $r_1=0.73$
suppresses corrections to scaling both in the case of the dynamic
structure functions as well as in static quantities,\cite{MKslab} and
(ii) that the deviations from scaling according to Figs.~\ref{S1pw3}
and \ref{S1pw10} have opposite signs depending on whether $r_1$ is
bigger or smaller than the optimal value $0.73$--- the corrections to
scaling which the finiteness of $c_0$ induces appear to play a major
role. 

\section{Summary and conclusions}\label{sec:concl}

We have presented a detailed study of the surface critical behavior of
isotropic Heisenberg ferromagnets at the ordinary transition, using
both sophisticated analytical tools as well as high-precision
Monte-Carlo spin dynamics simulations. In the \emph{analytical part}
of our work a continuum model that represents the corresponding
universality class of bulk and surface critical behavior---namely, an
appropriate semi-infinite extension of model J---has been formulated
and its field theory constructed. To this end we have determined the
relevant boundary contributions to the dynamic action functional which
are compatible with the general features (symmetries, detailed
balance, locality assumptions, conservation laws, etc.) of this class
of systems. These were shown to correspond to boundary conditions for
the resulting dynamic field theory.

In order to investigate the critical behavior of this semi-infinite
model J, we have employed two distinct RG schemes. The first is a
massless one based on the expansion about the upper critical dimension
$d_{\text{J}}^*=6$. To avoid the difficulties it has in handling the
problem adequately below the upper critical dimension $d^*=4$ of the
\emph{static} theory, we have designed and utilized an appropriate
extension of the massive RG scheme for bounded systems of Diehl and
Shpot.\cite{DS94,DS98} As usual, this works in fixed dimensions, and
avoids the dimensionality expansion. By combining the resulting RG
equations with the boundary operator expansion (\ref{eq:boephi}), we
have been able to obtain detailed predictions for the scaling behavior
of the surface structure function (\ref{eq:C11pomega}). The involved
critical exponents which govern power laws like (\ref{eq:C0}) are
related to known static bulk and surface critical exponents. In
particular, there is no independent new dynamic exponent associated
with the surface.\cite{rem:dynsurfexp}

We have checked our predictions by means of extensive Monte-Carlo spin
dynamics simulations. Our results depicted in Figs.~\ref{S10wfig} and
\ref{S1pw} corroborate the predicted dynamic scaling behavior. In
order to obtain such a good manifestation of scaling, we have found it
helpful and necessary to choose the ratio $r_1=J_1/J$ of the surface
and bulk exchange integrals $J_1$ and $J$ such that corrections to
scaling are suppressed. To achieve this objective we have optimized
the value of $r_1$ by requiring that the Monte Carlo data yield an
equilibrium order-parameter profile which satisfies a Dirichlet
boundary condition on the scale of the lattice constant in the sense
described in Sec.~\ref{sec:MC}.

According to our results displayed in
Figs.~\ref{S10wfig}--\ref{S1pw10}, this procedure is quite effective:
For $r_1=0.73$ and values inside a narrow range around this optimal
one, the simulation data for the dynamic structure functions exhibit
clear evidence of dynamic scaling in conformity with our predictions.
On the other hand, for values of $r_1$ outside this regime, the data
collapse is poor and the asymptotic behavior cannot be identified in a
convincing fashion. These observations are in conformity with those
made in a previous Monte Carlo investigation of the static surface
critical behavior by one of us.\cite{MKslab} However, in the study of
static quantities one is in a much better position because the scaling
regimes can be reached reasonably well even for non-optimal values of
$r_1$. This fact lends support to our belief that the dynamic scaling
behavior seen for $r_1=0.73$ can be trusted.

Finally, let us note that we have not taken into account any dipolar
forces in our study. To check our result by experiments (as should
become feasible in the near future thanks to facilities like the X-ray
free electron laser\cite{fel}), one must choose systems for which such
forces are negligible. Since even weak dipolar forces lead to the
formation of domains, one must also make sure that single domains are
investigated.

\begin{acknowledgments}
  We are grateful to K.\ Wiese for discussions during the
  beginning phase of this work. M.K.\ is indebted to the
  \emph{Institut f{\"u}r angewandte Informatik} at the Bergische
  Universit\"at Wuppertal for providing access to the parallel cluster
  ALiCE.  Partial support by DFG (for M.K.\ via the Heisenberg program
  under grant \# Kr 1322/2-1, for H.K.\ and H.W.D.\ via the Leibniz
  program Di 378/2-1 and SFB 237) is gratefully acknowledged.
\end{acknowledgments}

\appendix

\section{The irrelevance of energy conservation}
\label{sec:ec}

The dynamics of the lattice model we simulated, defined via the
equations of motion (\ref{eq:eqmot}), conserves the energy. Our aim
here is to show that this feature does not change the universality
class, which should therefore be represented by the semi-infinite
model J employed in our RG study. For the sake of simplicity, we
restrict ourselves to demonstrating the irrelevance of energy
conservation for the bulk case. The extension to the semi-infinite
case should be obvious.

Conservation of energy means that the energy density ${\mathcal
  E}(\bm{x},t)$ is a conserved density and hence a slow variable which
should be retained in a coarse-grained description on mesoscopic time
scales. Now, for vanishing mode-coupling constant $f_0$, model J
reduces to model B. How to modify the latter to account for energy
conservation is well known and leads to model D.\cite{HH77} We can
adapt the dynamics of model J along similar lines. The obvious result
is a modification of model J that differs from model D through the
addition of the former's mode-coupling terms. The Langevin equations
of this two-density model, which we call J', read
\begin{equation}\label{eq:stocheqJ1}
\dot{\bm{\phi}}(\bm{x},t) = \lambda_0\,{\left(\triangle\,
\frac{\delta {\mathcal{H}}'}{\delta {\bm{\phi}}}
+f_0\,\frac{\delta {\mathcal{H}}'}{\delta {\bm{\phi}}}
\times{\bm{\phi}}\right)}+\bm{\zeta}(\bm{x},t)
\end{equation}
and
\begin{equation}
  \label{eq:LeEnerg}
  \dot {\cal E}(\bm{x},t)=\lambda_0^{({\mathcal E})} \triangle\,
\frac{\delta {\mathcal H}'}{\delta {\cal E}}
+ \vartheta(\bm{x},t)\;, 
\end{equation}
where
\begin{equation}
\label{eq:HGL1}
{\mathcal H}' = {\int_{{\mathbb{R}}^d}}\!{\left[\frac{1}{2}\,
(\bm{\nabla} \bm{\phi})^2 + \frac{\tau_0}{2}\,\phi^2 +
\frac{u_0}{4!}\,|\bm{\phi}|^4 + \frac{1}{2}\,{\cal E}^2
+\frac{\gamma_0}{2} \,{\cal E}\,\phi^2 \right]}
\end{equation}
is the familiar Hamiltonian employed in the definition of models C, D,
and E (for the here considered case of an $n=3$ component order
parameter $\bm{\phi}$).  Both $\bm{\zeta}$ as well as $\vartheta$ are
Gaussian random forces with zero average; their variances are given by
Eq.~(\ref{eq:varzeta}) and
\begin{equation}\label{eq:varzeta1}
\left \langle \vartheta(\bm{x},t)\, \vartheta(\bm{x}',t')
\right\rangle =-2\lambda_0^{({\mathcal E})}\,
\triangle\,\delta(\bm{x}{-}\bm{x}')\,\delta(t{-}t')\;,
\end{equation}
respectively.

In the absence of coupling between the order parameter $\bm{\phi}$ and
the energy density ${\mathcal E}$, i.e., for $\gamma_0=0$, the dynamic
exponent of ${\mathcal E}$ takes its Gaussian value
\begin{equation}\label{eq:zE}
\mathfrak{z}_{\cal E} = 2\;,
\end{equation}
corresponding to ordinary diffusion. It is not difficult to see that
this result remains valid for $\gamma_0\ne 0$. For $d>4$, this follows
immediately from the observation that $\gamma_0$ is irrelevant in the
RG sense.

In studying the more interesting case $d\le 4$, we can benefit from
well-known results for the static theory described by the Hamiltonian
(\ref{eq:HGL1}), which is equivalent to the $\phi^4$-Hamiltonian
(\ref{eq:HGL}) except for a change $u_0\to U_0\equiv
u_0-3\,\gamma_0^2$ of the interaction constant (see, for example,
Refs.~\onlinecite{BdD75} and Ref.~\onlinecite{HH77}, and their
references). As a consequence, the corresponding static
renormalization functions can be expressed in terms of those of the
$\phi^4$-theory. In particular, the renormalized analog $U$ of $U_0$
can be introduced in analogy to Eq.~(\ref{eq:uren}) via
$U_0=Z_u(U)\,m^{4-d}\,U$, where $Z_u(u)$ is the renormalization
function of Sec.~\ref{sec:statsurfrfs}. Likewise, the renormalization
factor $Z_\gamma(U,\gamma^2)$ which relates
$\gamma_0=m^{(4-d)/2}\,Z_\gamma\,\gamma$ to its renormalized
counterpart $\gamma$ can be expressed in terms of known
renormalization functions of the $\phi^4$-theory. (It is a product of
$Z_{\phi^2}(U)$ and a factor linear in $\gamma^2$ whose $U$-dependent
coefficient derives from the additive counterterm that the
$\phi^4$-vertex function $\Gamma_{\phi^2\phi^2}$
requires.\cite{BdD75})

The resulting RG flow of $U$ and $\gamma$ has two nontrivial fixed
points at $U=u^*$: one at $\gamma^*=0$, and another one at
$(\gamma^*)^2=\text{const}\,{\alpha}/{\nu}$. The slopes
$\partial\beta_{\gamma^2} (U{=}u^*,\gamma^2)/\partial \gamma^2$ of the
beta function $\beta_{\gamma^2}\equiv \left.m\partial_m\right|_0\gamma^2$ at
these fixed points are given by $-\alpha/\nu$ and $\alpha/\nu$, respectively.

Since $\alpha<0$ in the three-component case we are concerned with
($\alpha\simeq -0.12$ for $d=3$, according to
Ref.~\onlinecite{GZJ98}), the infrared-stable fixed point is
$(U,\gamma)=(u^*,0)$. The results of Ref.~\onlinecite{BdD75} imply
that the running coupling constant associated with $\gamma$ tends to
zero as $m^{-\alpha/2\nu}$ in the limit $m\to 0$.  Thus the energy
density decouples asymptotically from the order parameter, so that the
result (\ref{eq:zE}) applies.

We can introduce the renormalized transport coefficient
$\lambda^{({\mathcal E})}$ via $\lambda_0^{({\mathcal
    E})}=m^{-2}\,Z_{\mathcal E}\,\lambda^{({\mathcal E})}$, where
$Z_{\mathcal E}$, the static renormalization factor of the energy
density, takes the value $1$ at the infrared-stable fixed point. The
ratio of transport coefficients $\lambda/\lambda^{({\mathcal E})}$ has
the asymptotic scale dependence $\sim m^{\mathfrak{z}-\mathfrak{z}_{\mathcal
    E}}$. If we substitute the values (\ref{eq:dynz}) and
(\ref{eq:zE}) for $\mathfrak{z}$ and $\mathfrak{z}_{\mathcal E}$, the
exponent becomes $\mathfrak{z}-\mathfrak{z}_{{\mathcal
    E}}=(d-2-\eta)/2$. Since this is positive in three dimensions, the
ratio approaches zero in the long-scale limit $m\to 0$. The upshot is
that the critical dynamics of the order parameter remains unaffected
by the coupling to the energy density, as claimed.


\begin{thebibliography}{59}
\expandafter\ifx\csname natexlab\endcsname\relax\def\natexlab#1{#1}\fi
\expandafter\ifx\csname bibnamefont\endcsname\relax
  \def\bibnamefont#1{#1}\fi
\expandafter\ifx\csname bibfnamefont\endcsname\relax
  \def\bibfnamefont#1{#1}\fi
\expandafter\ifx\csname citenamefont\endcsname\relax
  \def\citenamefont#1{#1}\fi
\expandafter\ifx\csname url\endcsname\relax
  \def\url#1{\texttt{#1}}\fi
\expandafter\ifx\csname urlprefix\endcsname\relax\def\urlprefix{URL }\fi
\providecommand{\bibinfo}[2]{#2}
\providecommand{\eprint}[2][]{\url{#2}}

\bibitem[{MEF()}]{MEF98} \bibinfo{note}{M. E. Fisher, Rev. Mod. Phys.
    {\bf 46}, 597 (1974); {\bf 70}, 653 (1998).}
  
\bibitem[{\citenamefont{Fisher}(1983)}]{Fis83}
  \bibinfo{author}{\bibfnamefont{M.~E.} \bibnamefont{Fisher}}, in
  \emph{\bibinfo{booktitle}{Critical Phenomena}}, edited by
  \bibinfo{editor}{\bibfnamefont{F.~J.~W.} \bibnamefont{Hahne}}
  (\bibinfo{publisher}{Springer-Verlag}, \bibinfo{address}{Berlin},
  \bibinfo{year}{1983}), vol. \bibinfo{volume}{186} of
  \emph{\bibinfo{series}{Lecture Notes in Physics}}, pp.
  \bibinfo{pages}{1--139}.
  
\bibitem[{\citenamefont{Halperin and Hohenberg}(1977)}]{HH77}
  \bibinfo{author}{\bibfnamefont{B.~I.} \bibnamefont{Halperin}}
  \bibnamefont{and} \bibinfo{author}{\bibfnamefont{P.~C.}
    \bibnamefont{Hohenberg}}, \bibinfo{journal}{Rev. Mod. Phys.}
  \textbf{\bibinfo{volume}{49}}, \bibinfo{pages}{435}
  (\bibinfo{year}{1977}).
  
\bibitem[{\citenamefont{Binder}(1983)}]{Bin83}
  \bibinfo{author}{\bibfnamefont{K.}~\bibnamefont{Binder}}, in
  \emph{\bibinfo{booktitle}{Phase Transitions and Critical
      Phenomena}}, edited by
  \bibinfo{editor}{\bibfnamefont{C.}~\bibnamefont{Domb}}
  \bibnamefont{and} \bibinfo{editor}{\bibfnamefont{J.~L.}
    \bibnamefont{Lebowitz}} (\bibinfo{publisher}{Academic},
  \bibinfo{address}{London}, \bibinfo{year}{1983}),
  vol.~\bibinfo{volume}{8}, pp. \bibinfo{pages}{1--144}.
  
\bibitem[{Die()}]{Diehl} \bibinfo{note}{H.~W. Diehl, in {\em Phase
      Transitions and Critical Phenomena}, edited by C. Domb and J.~L.
    Lebowitz (Academic Press, London, 1986), Vol.~10, pp.\ 75--267;
    Int.\ J.\ Mod.\ Phys.\ B {\bf 11}, 3503 (1997), preprint
    cond-mat/9610143.}
  
\bibitem[{Mod()}]{ModB_A} \bibinfo{note}{This happens, e.g., in the
    case of model B whose semi-infinite extensions B$_{\mathrm A}$ and
    B$_{\mathrm B}$, differing by the presence or absence of relevant
    nonconservative surface terms, represent distinct dynamic surface
    universality classes; see Refs.\ \onlinecite{DJ92},
    \onlinecite{Die94b}, and \onlinecite{WD95}.}
  
\bibitem[{\citenamefont{Diehl and Janssen}(1992)}]{DJ92}
  \bibinfo{author}{\bibfnamefont{H.~W.} \bibnamefont{Diehl}}
  \bibnamefont{and} \bibinfo{author}{\bibfnamefont{H.~K.}
    \bibnamefont{Janssen}}, \bibinfo{journal}{Phys.\ Rev.\ A}
  \textbf{\bibinfo{volume}{45}}, \bibinfo{pages}{7145}
  (\bibinfo{year}{1992}).
  
\bibitem[{\citenamefont{Diehl}(1994)}]{Die94b}
  \bibinfo{author}{\bibfnamefont{H.~W.} \bibnamefont{Diehl}},
  \bibinfo{journal}{Phys.\ Rev.\ B} \textbf{\bibinfo{volume}{49}},
  \bibinfo{pages}{2846} (\bibinfo{year}{1994}).
  
\bibitem[{\citenamefont{Wichmann and Diehl}(1995)}]{WD95}
  \bibinfo{author}{\bibfnamefont{F.}~\bibnamefont{Wichmann}}
  \bibnamefont{and} \bibinfo{author}{\bibfnamefont{H.~W.}
    \bibnamefont{Diehl}}, \bibinfo{journal}{Z.\ Phys.\ B}
  \textbf{\bibinfo{volume}{97}}, \bibinfo{pages}{251}
  (\bibinfo{year}{1995}).
  
\bibitem[{\citenamefont{Dietrich and Diehl}(1983)}]{DD83b}
  \bibinfo{author}{\bibfnamefont{S.}~\bibnamefont{Dietrich}}
  \bibnamefont{and} \bibinfo{author}{\bibfnamefont{H.~W.}
    \bibnamefont{Diehl}}, \bibinfo{journal}{Z.\ Phys.\ B}
  \textbf{\bibinfo{volume}{51}}, \bibinfo{pages}{343}
  (\bibinfo{year}{1983}), \bibinfo{note}{[Erratum {\bf 52}, 171
    (1983)]}.
  
\bibitem[{FD()}]{FD} \bibinfo{note}{D. Frank and V. Dohm, Phys.\ Rev.\ 
    Lett. {\bf 62}, 1864 (1989); Z. Phys. B {\bf 84}, 443 (1991).}
  
\bibitem[{XiG()}]{XiGo} \bibinfo{note}{G.~M. Xiong and C.~D. Gong, Z.
    Phys. B {\bf 74}, 379 (1989); J.  Phys. Condens. Matter {\bf 1},
    8673 (1989).}
  
\bibitem[{\citenamefont{Mail{\"a}nder
      et~al.}(1990)\citenamefont{Mail{\"a}nder, Dosch, Peisl, and
      Johnson}}]{MDPJ90}
  \bibinfo{author}{\bibfnamefont{L.}~\bibnamefont{Mail{\"a}nder}},
  \bibinfo{author}{\bibfnamefont{H.}~\bibnamefont{Dosch}},
  \bibinfo{author}{\bibfnamefont{J.}~\bibnamefont{Peisl}},
  \bibnamefont{and} \bibinfo{author}{\bibfnamefont{R.~L.}
    \bibnamefont{Johnson}}, \bibinfo{journal}{Phys.\ Rev.\ Lett.}
  \textbf{\bibinfo{volume}{64}}, \bibinfo{pages}{2527}
  (\bibinfo{year}{1990}).
  
\bibitem[{\citenamefont{Dosch}(1992)}]{Dos92}
  \bibinfo{author}{\bibfnamefont{H.}~\bibnamefont{Dosch}},
  \emph{\bibinfo{title}{Critical Phenomena at Surfaces and
      Interfaces}}, vol.  \bibinfo{volume}{126} of
  \emph{\bibinfo{series}{Springer Tracts in Modern Physics}}
  (\bibinfo{publisher}{Springer}, \bibinfo{address}{Berlin},
  \bibinfo{year}{1992}).
  
\bibitem[{\citenamefont{Krimmel et~al.}(1997)\citenamefont{Krimmel,
      Donner, Nickel, Dosch, Sutter, and Gr{\"u}bel}}]{KDN+97}
  \bibinfo{author}{\bibfnamefont{S.}~\bibnamefont{Krimmel}},
  \bibinfo{author}{\bibfnamefont{W.}~\bibnamefont{Donner}},
  \bibinfo{author}{\bibfnamefont{B.}~\bibnamefont{Nickel}},
  \bibinfo{author}{\bibfnamefont{H.}~\bibnamefont{Dosch}},
  \bibinfo{author}{\bibfnamefont{C.}~\bibnamefont{Sutter}},
  \bibnamefont{and}
  \bibinfo{author}{\bibfnamefont{G.}~\bibnamefont{Gr{\"u}bel}},
  \bibinfo{journal}{Phys.\ Rev.\ Lett.} \textbf{\bibinfo{volume}{78}},
  \bibinfo{pages}{3880} (\bibinfo{year}{1997}).
  
\bibitem[{Xra()}]{Xraytheory} \bibinfo{note}{For a detailed exposition
    of the theoretical background on surface X-ray and neutron
    scattering, see S. Dietrich and H. Wagner, Phys.\ Rev.\ Lett. {\bf
      51}, 1469 (1983); Z.\ Phys.\ B {\bf 56}, 207 (1984);
    \emph{ibida} {\bf 59}, 35 (1985); S. Dietrich and A. Haase, Phys.\ 
    Rep. {\bf 260}, 1 (1995).}
  
\bibitem[{fel()}]{fel} \bibinfo{note}{TESLA Technical Design Report,
    Part V, DESY, March 2001, http://tesla.desy.de/tdr/.}
  
\bibitem[{\citenamefont{Krech et~al.}(2001)\citenamefont{Krech, Karl,
      and Diehl}}]{KKD01}
  \bibinfo{author}{\bibfnamefont{M.}~\bibnamefont{Krech}},
  \bibinfo{author}{\bibfnamefont{H.}~\bibnamefont{Karl}},
  \bibnamefont{and} \bibinfo{author}{\bibfnamefont{H.~W.}
    \bibnamefont{Diehl}}, \bibinfo{journal}{Physica A}
  \textbf{\bibinfo{volume}{297}}, \bibinfo{pages}{64}
  (\bibinfo{year}{2001}), \bibinfo{note}{cond-mat/0102131}.
  
\bibitem[{\citenamefont{Ma and Mazenko}(1975)}]{MM75}
  \bibinfo{author}{\bibfnamefont{S.}~\bibnamefont{Ma}}
  \bibnamefont{and} \bibinfo{author}{\bibfnamefont{G.~F.}
    \bibnamefont{Mazenko}}, \bibinfo{journal}{Phys.\ Rev.\ B}
  \textbf{\bibinfo{volume}{11}}, \bibinfo{pages}{4077}
  (\bibinfo{year}{1975}).
  
\bibitem[{\citenamefont{Kawasaki}(1975)}]{Kaw75}
  \bibinfo{author}{\bibfnamefont{K.}~\bibnamefont{Kawasaki}},
  \bibinfo{journal}{Prog. Theor. Phys.} \textbf{\bibinfo{volume}{54}},
  \bibinfo{pages}{1665} (\bibinfo{year}{1975}).
  
\bibitem[{\citenamefont{Janssen}(1976)}]{Jan76}
  \bibinfo{author}{\bibfnamefont{H.-K.} \bibnamefont{Janssen}},
  \bibinfo{journal}{Z.\ Phys.\ B} \textbf{\bibinfo{volume}{23}},
  \bibinfo{pages}{377} (\bibinfo{year}{1976}).
  
\bibitem[{\citenamefont{Bausch et~al.}(1976)\citenamefont{Bausch,
      Janssen, and Wagner}}]{BJW76}
  \bibinfo{author}{\bibfnamefont{R.}~\bibnamefont{Bausch}},
  \bibinfo{author}{\bibfnamefont{H.~K.} \bibnamefont{Janssen}},
  \bibnamefont{and}
  \bibinfo{author}{\bibfnamefont{H.}~\bibnamefont{Wagner}},
  \bibinfo{journal}{Z. Phys. B} \textbf{\bibinfo{volume}{24}},
  \bibinfo{pages}{113} (\bibinfo{year}{1976}).
  
\bibitem[{\citenamefont{Dohm}(1976)}]{Doh76}
  \bibinfo{author}{\bibfnamefont{V.}~\bibnamefont{Dohm}},
  \bibinfo{journal}{Solid State Comm.} \textbf{\bibinfo{volume}{20}},
  \bibinfo{pages}{657} (\bibinfo{year}{1976}).
  
\bibitem[{\citenamefont{Janssen}(1979)}]{Jan79}
  \bibinfo{author}{\bibfnamefont{H.~K.} \bibnamefont{Janssen}}, in
  \emph{\bibinfo{booktitle}{Dynamical Critical Phenomena and Related
      Topics}}, edited by \bibinfo{editor}{\bibfnamefont{C.~P.}
    \bibnamefont{Enz}} (\bibinfo{publisher}{Springer-Verlag},
  \bibinfo{address}{Berlin}, \bibinfo{year}{1979}), pp.
  \bibinfo{pages}{25--47}.
  
\bibitem[{\citenamefont{Diehl and Shpot}(1994)}]{DS94}
  \bibinfo{author}{\bibfnamefont{H.~W.} \bibnamefont{Diehl}}
  \bibnamefont{and}
  \bibinfo{author}{\bibfnamefont{M.}~\bibnamefont{Shpot}},
  \bibinfo{journal}{Phys.\ Rev.\ Lett.} \textbf{\bibinfo{volume}{73}},
  \bibinfo{pages}{3431} (\bibinfo{year}{1994}).
  
\bibitem[{\citenamefont{Diehl and Shpot}(1998)}]{DS98}
  \bibinfo{author}{\bibfnamefont{H.~W.} \bibnamefont{Diehl}}
  \bibnamefont{and}
  \bibinfo{author}{\bibfnamefont{M.}~\bibnamefont{Shpot}},
  \bibinfo{journal}{Nucl. Phys. B} \textbf{\bibinfo{volume}{528}},
  \bibinfo{pages}{595} (\bibinfo{year}{1998}),
  \bibinfo{note}{cond-mat/9804083}.
  
\bibitem[{\citenamefont{Dietrich and Diehl}(1981)}]{DD81c}
  \bibinfo{author}{\bibfnamefont{S.}~\bibnamefont{Dietrich}}
  \bibnamefont{and} \bibinfo{author}{\bibfnamefont{H.~W.}
    \bibnamefont{Diehl}}, \bibinfo{journal}{Z.\ Phys.\ B}
  \textbf{\bibinfo{volume}{43}}, \bibinfo{pages}{315}
  (\bibinfo{year}{1981}).
  
\bibitem[{\citenamefont{McAvity and Osborn}(1993)}]{MO93}
  \bibinfo{author}{\bibfnamefont{D.~M.} \bibnamefont{McAvity}}
  \bibnamefont{and}
  \bibinfo{author}{\bibfnamefont{H.}~\bibnamefont{Osborn}},
  \bibinfo{journal}{Nucl.\ Phys.\ B} \textbf{\bibinfo{volume}{406}},
  \bibinfo{pages}{655} (\bibinfo{year}{1993}).
  
\bibitem[{SD()}]{SD} \bibinfo{note}{M. Krech, A. Bunker, and D.~P.
    Landau, Comp. Phys. Commun. 111, 1 (1998); D.~P. Landau, S.-H.
    Tsai, M. Krech, and A. Bunker, Int. J. Mod.  Phys. {\bf 10}, 1541
    (1999); D.~P. Landau, A. Bunker, H.~G. Evertz, M. Krech, and S.-H.
    Tsai, Prog. Theor. Phys. Suppl. {\bf 138}, 423 (2000).}
  
\bibitem[{DPL()}]{DPLMK} \bibinfo{note}{D. P. Landau and M. Krech, J.
    Phys. Cond. Mat. {\bf 11}, R179 (1999); M. Krech and D. P. Landau,
    Phys. Rev. B {\bf 60}, 3375 (1999).}
  
\bibitem[{\citenamefont{Krech}(2000)}]{MKslab}
  \bibinfo{author}{\bibfnamefont{M.}~\bibnamefont{Krech}},
  \bibinfo{journal}{Phys.\ Rev.\ B} \textbf{\bibinfo{volume}{62}},
  \bibinfo{pages}{6360} (\bibinfo{year}{2000}),
  \bibinfo{note}{cond-mat/0006448, and references therein.}
  
\bibitem[{\citenamefont{Janssen}(1992)}]{Jan92}
  \bibinfo{author}{\bibfnamefont{H.~K.} \bibnamefont{Janssen}}, in
  \emph{\bibinfo{booktitle}{From Phase Transitions to Chaos}}, edited
  by \bibinfo{editor}{\bibfnamefont{G.}~\bibnamefont{Gy{\"o}rgyi}},
  \bibinfo{editor}{\bibfnamefont{I.}~\bibnamefont{Kondor}},
  \bibinfo{editor}{\bibfnamefont{L.}~\bibnamefont{Sasv{\'a}ri}},
  \bibnamefont{and}
  \bibinfo{editor}{\bibfnamefont{T.}~\bibnamefont{Tel}}
  (\bibinfo{publisher}{World Scientific},
  \bibinfo{address}{Singapore}, \bibinfo{year}{1992}), pp.
  \bibinfo{pages}{68--91}.
  
\bibitem[{\citenamefont{de~Dominicis}(1976)}]{dDom76}
  \bibinfo{author}{\bibfnamefont{C.}~\bibnamefont{de~Dominicis}},
  \bibinfo{journal}{J. Phys. (Paris) Colloq.}
  \textbf{\bibinfo{volume}{37}}, \bibinfo{pages}{C1}
  (\bibinfo{year}{1976}).
  
\bibitem[{\citenamefont{Martin et~al.}(1973)\citenamefont{Martin,
      Siggia, and Rose}}]{MSR73} \bibinfo{author}{\bibfnamefont{P.~C.}
    \bibnamefont{Martin}}, \bibinfo{author}{\bibfnamefont{E.~D.}
    \bibnamefont{Siggia}}, \bibnamefont{and}
  \bibinfo{author}{\bibfnamefont{A.}~\bibnamefont{Rose}},
  \bibinfo{journal}{Phys. Rev. A} \textbf{\bibinfo{volume}{8}},
  \bibinfo{pages}{423} (\bibinfo{year}{1973}).
  
\bibitem[{\citenamefont{Karl}(2000)}]{Kar00}
  \bibinfo{author}{\bibfnamefont{H.}~\bibnamefont{Karl}},
  \emph{\bibinfo{title}{{Z}um {E}in\-flu{\ss} von
      {O}ber\-fl{\"a}\-chen auf das dy\-na\-mi\-sche kri\-ti\-sche
      {V}er\-hal\-ten von iso\-tro\-pen
      {H}ei\-sen\-berg-{F}er\-ro\-mag\-ne\-ten}}, Dissertation, U.\ 
  Essen (\bibinfo{publisher}{Shaker-Verlag},
  \bibinfo{address}{Aachen}, \bibinfo{year}{2000}).
  
\bibitem[{\citenamefont{Amit and Peliti}(1982)}]{AP82}
  \bibinfo{author}{\bibfnamefont{D.~J.} \bibnamefont{Amit}}
  \bibnamefont{and}
  \bibinfo{author}{\bibfnamefont{L.}~\bibnamefont{Peliti}},
  \bibinfo{journal}{Ann. Phys. (USA)} \textbf{\bibinfo{volume}{140}},
  \bibinfo{pages}{207} (\bibinfo{year}{1982}).
  
\bibitem[{BJW()}]{BJWcomins} \bibinfo{note}{The essence of this
    argument is nothing but a combination of the way in which the
    fluctuation-dissipation theorem was exploited in Ref.\ 
    \onlinecite{BJW76} with the consequences implied by the minimal
    subtraction of poles. The discussion of the renormalization of
    insertions of the composite operator
    $\tilde{\bm{\phi}}\times\bm{\phi}$ given in this reference is
    restricted to \emph{single} insertions. However, from our
    reasoning yielding the possible $\bm{K}$-dependent counterterms it
    is clear that the cumulants with arbitrarily many insertions,
    derived from the functional (\ref{eq:Wren}), must be uv finite as
    well.}
  
\bibitem[{com()}]{comVF} \bibinfo{note}{In the bulk case, this vertex
    function, defined algebraically in the conventional way, simply
    turns out to be the operator inverse of
    $\langle\phi^\alpha\tilde{\phi}^\beta\rangle^{\text{cum}}$. In
    cases with boundaries as considered here, already the inverse of
    the free response propagator is uniquely defined only when
    considered as acting on a space of functions that satisfy the
    boundary conditions imposed on the free response propagator. The
    inverse then simply is given by
    $\delta^2/\delta\tilde{\phi}^\alpha\delta\phi^\beta$ of the
    Gaussian part of the action ${\mathcal{J}}$. The full
    $\Gamma_{\tilde{\phi}^\alpha\phi^\beta}$ may be considered to be
    defined graphically as being given by this zero-loop term minus
    the sum of all one-particle irreducible graphs with one external
    $\tilde{\phi}$ and one external $\phi$ leg.}
  
\bibitem[{HKJ()}]{HKJlec} \bibinfo{note}{This result may be gleaned
    from H.~K.\ Janssen's lecture notes (unpublished) on his course on
    critical dynamics.}
  
\bibitem[{\citenamefont{Wagner}(1970)}]{Wag70}
  \bibinfo{author}{\bibfnamefont{H.}~\bibnamefont{Wagner}},
  \bibinfo{journal}{Phys. Lett.} \textbf{\bibinfo{volume}{33A}},
  \bibinfo{pages}{58} (\bibinfo{year}{1970}).
  
\bibitem[{spw()}]{spwavefreq} \bibinfo{note}{The result
    (\ref{eq:dynz}) can also be derived for $T<T_{\text{c}}$ via a
    simple scaling argument:\cite{MM75,HH77} According to
    hydrodynamics the long-wave-length spin waves must have a
    frequency $\omega_q\sim M\,q^2$, where $M$ is the (spontaneous)
    magnetization. Now $M\sim \xi^{-\beta/\nu}$ as $T\to
    T_{\text{c}}-$. For $2<d<4$, where the hyperscaling law
    $\beta/\nu=(d-2+\eta)/2$ holds, we thus have $\omega_q\sim
    \xi^{-(d+2+\eta)/2}\,(q\xi)^2$. However, for $4<d<6$, hyperscaling
    is broken, $\beta/\nu$ takes its mean-field value $1$, one must
    take into account that $M\sim \xi^{-(d-2)/2}/\sqrt{\bar{u}(\xi)}$,
    where $\bar{u}(\xi)\sim u\,\xi^{4-d}$ is the running $\phi^4$
    coupling constant. The resulting scaling form
    $\omega_q=q^{1+d/2}\,\Omega(q\xi,u\xi^{4-d})$ reflects the
    dangerous character of the irrelevant variable $u$ and the
    concomitant necessity of including a second length (`thermodynamic
    length') besides $\xi$ in the scaling.}
  
\bibitem[{rem({\natexlab{a}})}]{rem:normal} \bibinfo{note}{We do not
    consider the possibility that the $O(3)$ symmetry of the
    Hamiltonian is broken by surface terms such as a surface magnetic
    field or surface spin anisotropies. In the former case, a
    normal\cite{Diehl} surface transition would be possible. Likewise,
    an easy-axis surface spin anisotropy could give rise to an
    anisotropic special transition\cite{DE84} in three dimensions if
    the corresponding surface interaction constant is sufficiently
    enhanced.}
  
\bibitem[{\citenamefont{Diehl and Eisenriegler}(1984)}]{DE84}
  \bibinfo{author}{\bibfnamefont{H.~W.} \bibnamefont{Diehl}}
  \bibnamefont{and}
  \bibinfo{author}{\bibfnamefont{E.}~\bibnamefont{Eisenriegler}},
  \bibinfo{journal}{Phys.\ Rev.\ B} \textbf{\bibinfo{volume}{30}},
  \bibinfo{pages}{300} (\bibinfo{year}{1984}).
  
\bibitem[{\citenamefont{Diehl and Dietrich}(1980)}]{DD80}
  \bibinfo{author}{\bibfnamefont{H.~W.} \bibnamefont{Diehl}}
  \bibnamefont{and}
  \bibinfo{author}{\bibfnamefont{S.}~\bibnamefont{Dietrich}},
  \bibinfo{journal}{Phys.\ Lett.} \textbf{\bibinfo{volume}{80A}},
  \bibinfo{pages}{408} (\bibinfo{year}{1980}).
  
\bibitem[{\citenamefont{Diehl and Dietrich}(1981)}]{DD81a}
  \bibinfo{author}{\bibfnamefont{H.~W.} \bibnamefont{Diehl}}
  \bibnamefont{and}
  \bibinfo{author}{\bibfnamefont{S.}~\bibnamefont{Dietrich}},
  \bibinfo{journal}{Z.\ Phys.\ B} \textbf{\bibinfo{volume}{42}},
  \bibinfo{pages}{65} (\bibinfo{year}{1981}), \bibinfo{note}{erratum:
    {\bf 43}, 281 (1981)}.
  
\bibitem[{\citenamefont{Symanzik}(1981)}]{Sym81}
  \bibinfo{author}{\bibfnamefont{K.}~\bibnamefont{Symanzik}},
  \bibinfo{journal}{Nucl.\ Phys.\ B} \textbf{\bibinfo{volume}{190}},
  \bibinfo{pages}{1} (\bibinfo{year}{1981}).
  
\bibitem[{\citenamefont{Krech et~al.}(1995)\citenamefont{Krech,
      Eisenriegler, and Dietrich}}]{KED95}
  \bibinfo{author}{\bibfnamefont{M.}~\bibnamefont{Krech}},
  \bibinfo{author}{\bibfnamefont{E.}~\bibnamefont{Eisenriegler}},
  \bibnamefont{and}
  \bibinfo{author}{\bibfnamefont{S.}~\bibnamefont{Dietrich}},
  \bibinfo{journal}{Phys.\ Rev.\ E} \textbf{\bibinfo{volume}{52}},
  \bibinfo{pages}{1345} (\bibinfo{year}{1995}).
  
\bibitem[{rem({\natexlab{b}})}]{rem:rense} \bibinfo{note}{A definition
    of this renormalized surface enhancement $c$ can be found in 3.2
    of Ref.~\onlinecite{DS98}, but will not be needed in the sequel.}
  
\bibitem[{rem({\natexlab{c}})}]{rem:opcons}
  \bibinfo{note}{Conservation of the order parameter implies that the
    zero-momentum contribution
    $\tilde{\bm{\phi}}_{\bm{q}{=}\bm{0}}(t)\cdot
    \dot{\bm{\phi}}_{\bm{q}{=}\bm{0}}$ to the bulk action must retain
    its form upon renormalization, so that the trivial equation of
    motion $\dot{\bm{\phi}}_{\bm{q}{=}\bm{0}}=\partial_t\int
    d^dx\,\bm{\phi}=\bm{0}$ remains valid.}
  
\bibitem[{\citenamefont{Bagnuls and Bervillier}(1981)}]{BB81}
  \bibinfo{author}{\bibfnamefont{C.}~\bibnamefont{Bagnuls}}
  \bibnamefont{and}
  \bibinfo{author}{\bibfnamefont{C.}~\bibnamefont{Bervillier}},
  \bibinfo{journal}{Phys. Rev. B} \textbf{\bibinfo{volume}{24}},
  \bibinfo{pages}{1226} (\bibinfo{year}{1981}).
  
\bibitem[{\citenamefont{Zinn-Justin}(1996)}]{ZJ96}
  \bibinfo{author}{\bibfnamefont{J.}~\bibnamefont{Zinn-Justin}},
  \emph{\bibinfo{title}{Quantum Field Theory and Critical Phenomena}},
  International series of monographs on physics
  (\bibinfo{publisher}{Clarendon Press}, \bibinfo{address}{Oxford},
  \bibinfo{year}{1996}), \bibinfo{edition}{3rd} ed.
  
\bibitem[{\citenamefont{Wolff}(1989)}]{Wol89}
  \bibinfo{author}{\bibfnamefont{U.}~\bibnamefont{Wolff}},
  \bibinfo{journal}{Phys. Rev. Lett.} \textbf{\bibinfo{volume}{62}},
  \bibinfo{pages}{361} (\bibinfo{year}{1989}).
  
\bibitem[{\citenamefont{Chen et~al.}(1993)\citenamefont{Chen,
      Ferrenberg, and Landau}}]{CFL93}
  \bibinfo{author}{\bibfnamefont{K.}~\bibnamefont{Chen}},
  \bibinfo{author}{\bibfnamefont{A.~M.} \bibnamefont{Ferrenberg}},
  \bibnamefont{and} \bibinfo{author}{\bibfnamefont{D.~P.}
    \bibnamefont{Landau}}, \bibinfo{journal}{Phys. Rev. B}
  \textbf{\bibinfo{volume}{48}}, \bibinfo{pages}{3249}
  (\bibinfo{year}{1993}).
  
\bibitem[{clu()}]{cluerr} \bibinfo{note}{A.~M. Ferrenberg, D.~P.
    Landau, and Y.~J. Wong, Phys. Rev. Lett.  {\bf 69}, 3382 (1993);
    L.~N. Shchur and H.~W.~J. Bl\"ote, Phys. Rev. E {\bf 55}, R4905
    (1997).}
  
\bibitem[{\citenamefont{Ferrenberg and Landau}()}]{AFMDPL}
  \bibinfo{author}{\bibfnamefont{A.~M.} \bibnamefont{Ferrenberg}}
  \bibnamefont{and} \bibinfo{author}{\bibfnamefont{D.~P.}
    \bibnamefont{Landau}}, \bibinfo{note}{private communication}.
  
\bibitem[{\citenamefont{Guida and Zinn-Justin}(1998)}]{GZJ98}
  \bibinfo{author}{\bibfnamefont{R.}~\bibnamefont{Guida}}
  \bibnamefont{and}
  \bibinfo{author}{\bibfnamefont{J.}~\bibnamefont{Zinn-Justin}},
  \bibinfo{journal}{J. Phys. A} \textbf{\bibinfo{volume}{31}},
  \bibinfo{pages}{8103} (\bibinfo{year}{1998}).
  
\bibitem[{rem({\natexlab{d}})}]{rem:omegaf} \bibinfo{note}{The
    dimensionality expansion of its counterpart for $4\le d <6$,
    namely $\omega_f\equiv\beta_f'(0,f^*)$, is known to first order:
    According to Ref.~\onlinecite{BJW76}, it reads $\omega_f=6-d
    +O{[(6-d)^2]}$.}
  
\bibitem[{rem({\natexlab{e}})}]{rem:dynsurfexp} \bibinfo{note}{If one
    were naive, one might expect that the analog $(d+2-\eta_\|)/2$ of
    the dynamic exponent $\mathfrak{z}=(d+2-\eta)/2$ might play the
    role of the latter in surface quantities. However, this is not the
    case.}
  
\bibitem[{\citenamefont{Br\'ezin and de~Dominicis}(1975)}]{BdD75}
  \bibinfo{author}{\bibfnamefont{E.}~\bibnamefont{Br\'ezin}}
  \bibnamefont{and}
  \bibinfo{author}{\bibfnamefont{C.}~\bibnamefont{de~Dominicis}},
  \bibinfo{journal}{Phys. Rev. B} \textbf{\bibinfo{volume}{12}},
  \bibinfo{pages}{4954} (\bibinfo{year}{1975}).

\end{thebibliography}
\end{document}